\documentclass[fleqn,usenatbib]{mnras}

\usepackage{newtxtext,newtxmath}
\usepackage[T1]{fontenc}

\usepackage{graphicx}
\usepackage{amsmath}
\usepackage{hyperref}
\usepackage{bm}
\usepackage[dvipsnames, table]{xcolor}
\usepackage{tikz}
\usepackage[normalem]{ulem}
\usepackage{orcidlink}

\newcommand{\hMpc}{\,h\text{Mpc}^{-1}}
\newcommand{\Mpch}{\,h^{-1}\text{Mpc}}
\newcommand{\Gpch}{\,h^{-1}\text{Gpc}}
\newcommand{\Omegam}{\Omega_{\rm m}}
\newcommand{\Omegab}{\Omega_{\rm b}}
\newcommand{\omegac}{\omega_{\rm cdm}}
\newcommand{\omegab}{\omega_{\rm b}}

\newcommand{\Neff}{N_{\rm eff}}
\newcommand{\Msunh}{\,h^{-1}{\rm M_{\odot}}}
\newcommand{\Q}{{\rm Q}}
\newcommand{\nrun}{{\rm d}n_s/{\rm d}\ln k}
\newcommand{\gauss}{\mathcal{N}}    
\newcommand{\unif}{\mathcal{U}}

\usepackage{array}

\usepackage{fontawesome}

\title[Density-split clustering in BOSS]{Cosmological constraints from density-split clustering in the BOSS CMASS galaxy sample}

\author[]{\parbox{\textwidth}{
Enrique Paillas$^{1, 2}$\thanks{E-mail: enrique.paillas@uwaterloo.ca}\orcidlink{0000-0002-4637-2868},
Carolina Cuesta-Lazaro$^{3, 4, 5}$\orcidlink{0000-0002-6069-2999},
Will J. Percival$^{1, 2, 6}$\orcidlink{0000-0002-0644-5727},
Seshadri Nadathur$^{7}$\orcidlink{0000-0001-9070-3102},
\mbox{Yan-Chuan Cai}$^{8}$\orcidlink{0000-0002-2128-866X},
\mbox{Sihan Yuan}$^{9, 10, 11}$\orcidlink{0000-0002-5992-7586},
Florian Beutler$^{8}$\orcidlink{0000-0003-0467-5438},
Arnaud de Mattia$^{12}$\orcidlink{0000-0003-0920-2947},
Daniel Eisenstein$^{3}$\orcidlink{0000-0002-2929-3121},
Daniel Forero-Sanchez$^{13}$\orcidlink{0000-0001-5957-332X},
Nelson Padilla$^{14}$\orcidlink{0000-0001-9850-9419},
Mathilde Pinon$^{12}$\orcidlink{0009-0009-3228-7126},
Vanina Ruhlmann-Kleider$^{12}$\orcidlink{0009-0000-6063-6121},
Ariel~G.~S\'anchez$^{15}$\orcidlink{0000-0003-1198-831X},
Georgios Valogiannis$^{16,17}$\orcidlink{0000-0003-0805-1470},
and Pauline Zarrouk$^{18}$\orcidlink{0000-0002-7305-9578}
}
\vspace*{4pt} \\
\scriptsize $^{1}$ Waterloo Centre for Astrophysics, University of Waterloo, Waterloo, ON N2L 3G1, Canada \\
\scriptsize $^{2}$ Department of Physics and Astronomy, University of Waterloo, 
Waterloo, ON N2L 3G1, Canada \\
\scriptsize $^{3}$ Center for Astrophysics | Harvard \& Smithsonian, 60 Garden St, Cambridge, MA 02138, USA  \\
\scriptsize $^{4}$ The NSF AI Institute for Artificial Intelligence and Fundamental Interactions \\
\scriptsize $^{5}$ Department of Physics, Massachusetts Institute of Technology, Cambridge, MA 02139, USA \\
\scriptsize $^{6}$ Perimeter Institute for Theoretical Physics, 31 Caroline St North, Waterloo, ON N2L 2Y5, Canada \\
\scriptsize $^{7}$ Institute of Cosmology and Gravitation, University of Portsmouth, Burnaby Road, Portsmouth, PO1 3FX, UK \\
\scriptsize $^{8}$ Institute for Astronomy, University of Edinburgh, Blackford Hill, Edinburgh, EH9 3HJ, UK\\
\scriptsize $^{9}$ Kavli Institute for Particle Astrophysics and Cosmology, Stanford University, 452 Lomita Mall, Stanford, CA 94305, USA\\
\scriptsize $^{10}$ Department of Physics, Stanford University, 382 Via Pueblo Mall, Stanford, CA 94305, USA\\
\scriptsize $^{11}$ SLAC National Accelerator Laboratory, 2575 Sand Hill Road, Menlo Park, CA  94025, USA \\
\scriptsize $^{12}$ IRFU, CEA, Universit\'e Paris-Saclay, F-91191 Gif-sur-Yvette, France  \\
\scriptsize $^{13}$ Institute of Physics, Laboratory of Astrophysics, \'Ecole Polytechnique F\'ed\'erale de Lausanne (EPFL), Observatoire de Sauverny, CH-1290 Versoix, Switzerland \\
\scriptsize $^{14}$ Instituto de Astronom\'ia Te\'orica y Experimental (IATE), CONICET-Universidad Nacional de C\'ordoba, Laprida 854, X5000BGR, C\'ordoba, Argentina \\
\scriptsize $^{15}$ Max-Planck-Institut f\"ur Extraterrestrische Physik, Postfach 1312, Giessenbachstr., D-85748 Garching, Germany \\
\scriptsize $^{16}$ Department of Physics, Harvard University, Cambridge, MA, 02138, USA \\
\scriptsize $^{17}$ Department of Astronomy and Astrophysics, University of Chicago, Chicago, IL, 60637, USA \\
\scriptsize $^{18}$ Sorbonne Universit\'e, Universit\'e Paris Diderot, Sorbonne Paris Cit\'e, CNRS,
Laboratoire de Physique Nucl\'eaire et de Hautes Energies (LPNHE), 4 place Jussieu, F-75252, Paris
Cedex 5, France
\vspace*{-2pt}
}

\pubyear{2022}

\begin{document}
\label{firstpage}
\pagerange{\pageref{firstpage}--\pageref{lastpage}}
\maketitle

\begin{abstract}
We present a clustering analysis of the BOSS DR12 CMASS galaxy sample, combining measurements of the galaxy two-point correlation function and density-split clustering down to a scale of $1 \Mpch$. Our theoretical framework is based on emulators trained on high-fidelity mock galaxy catalogues that forward model the cosmological dependence of the clustering statistics within an extended-$\Lambda$CDM framework, including redshift-space and Alcock-Paczynski distortions. Our base-$\Lambda$CDM analysis finds $\omegac = 0.1201\pm 0.0022$, $\sigma_8 = 0.792\pm 0.034$, and $n_s = 0.970\pm 0.018$, corresponding to $f\sigma_8 = 0.462\pm 0.020$ at $z \approx 0.525$, which is in agreement with Planck 2018 predictions and various clustering studies in the literature. We test single-parameter extensions to base-$\Lambda$CDM, varying the running of the spectral index, the dark energy equation of state, and the density of massless relic neutrinos, finding no compelling evidence for deviations from the base model. We model the galaxy-halo connection using a halo occupation distribution framework, finding signatures of environment-based assembly bias in the data. We validate our pipeline against mock catalogues that match the clustering and selection properties of CMASS, showing that we can recover unbiased cosmological constraints even with a volume 84 times larger than the one used in this study.
\end{abstract}


\section{Introduction}

Within the vast cosmic structures we observe today, signatures of primordial features are intermixed with non-linear processes that shape the evolution of galaxies, constituting a ground that is challenging to model but rich in information. In our standard cosmological model, the $\Lambda$CDM paradigm, the present-day Universe is the result of a hierarchical structure formation scenario that started from primordial density perturbations, which were amplified during inflation and continued to grow through gravitational collapse until today \citep{Guth1982, Hawking1982}. Galaxy clustering has been pivotal in testing this hypothesis from late-time Universe data by characterizing the way in which matter is distributed in space, using galaxies as biased tracers of the underlying dark matter distribution.

The most common way to approach galaxy clustering is through the two-point correlation function, or its Fourier pair, the power spectrum. These statistics provide a nearly-complete description of the galaxy density field on large scales and encode information about physics of the early Universe \citep{Peebles1980}. Baryon acoustic oscillations (BAO), which originate from sound waves in the photon-baryon plasma before recombination \citep{Peebles1970, Sunyaev1970}, leave an imprint on the matter distribution, which is detected as a bump in the correlation function or as wiggles in the power spectrum \citep{Eisenstein1998:astro-ph/9709112}. This acoustic feature works as a standard ruler, allowing us to measure the expansion rate of the Universe at different epochs \citep{Percival2001:astro-ph/0105252, Eisenstein2005, Cole2005:astro-ph/0501174, eboss2020, Moon2023:2304.08427}. In addition to BAO, the full shape of the galaxy correlation function and power spectrum contains an enormous wealth of information due to its sensitivity to the growth rate of cosmic structure \citep{Blake2011:1104.2948, Reid2012:1203.6641, Alam2017:1607.03155, Brieden2021:2106.07641}, dark energy, the physics of neutrinos \citep{Zhang2022:2111.05739}, primordial non-Gaussianity \citep{Moradinezhad2021:2010.14523}, and the galaxy-halo connection \citep{Yuan2022:2110.11412}.

Models based on perturbation theory provide an accurate description of galaxy clustering data on linear and mildly non-linear scales, allowing the extraction of information from the full shape of the power spectrum \citep[e.g.,][]{Sanchez2016, Grieb2017:1607.03143, d'Amico2020:1909.05271, Troster2020:1909.11006, Bautista2021, Philcox2021a:2112.04515, Semenaite2022:2111.0315, Semenaite2023:2210.07304}. The assumptions behind these models tend to break down in the highly non-linear regime ($\lesssim 20 \Mpch$), which has motivated the development of models calibrated on N-body simulations that can accurately describe the galaxy field on scales where non-linear physical processes become relevant. This has recovered information from a portion of the survey data that is usually discarded from the standard clustering analyses, providing clues not only about cosmology but also in regards to galaxy evolution and its connection to the dark matter halo field \citep{Kobayashi2020, Zhai2023:2203.08999, Chapman2022:2106.14961, Lange2022:2101.12261, Lange2023:2301.08692, Yuan2022:2110.11412}. 

Two-point functions provide a complete description of Gaussian density fields. This is satisfied at large scales, where non-linear evolution is mild, and the Gaussianity of the primordial fluctuations are still largely preserved. The late-time Universe, however, is highly non-Gaussian at small scales due to non-linear evolution, and higher-order summary statistics are required to capture all the information. Thanks to substantial theoretical and algorithmic development over the last years, measurements of N-point correlation functions \citep{Slepian2017a:1512.02231, Philcox2021b:2108.01670, Sugiyama2023:2302.06808} and polyspectra \citep{Gil-Marin2017:1606.00439, Gualdi2021, Philcox2021a:2112.04515} are now being performed on data, which not only tightens the constraining power on $\Lambda$CDM, but also opens an avenue for testing potential signatures of physics beyond our fiducial model, such as parity violation \citep{Philcox2022:2206.04227, Hou2023a:2206.03625}. However, the measurement of these N-point statistics remains challenging due to their high computational demands and the large volumes that are needed to detect them with enough statistical significance. This has motivated the development of robust, informative, and efficient clustering methods that can access the information that leaks into higher orders, which can then be complemented and cross-validated with the standard N-point clustering analysis.

Several alternative summary statistics that meet these criteria have been proposed in the literature, including k-th nearest neighbor statistics \citep{Banerjee2021}, wavelet scattering transforms \citep{Valogiannis2021, Valogiannis2022:2204.13717}, void statistics \citep{Lavaux2012:1110.0345}, marked correlations \citep{White2016:1609.08632, Massara2022}, skew spectra \citep{Hou2023b:2210.12743}, and Minkowski functionals \citep{Lippich2021:2012.08529}. Among these novel methods, \cite{Paillas2021} proposed to perform a clustering analysis split by local density, combining the information content of different environments of the cosmic web. \cite{Paillas2021} showed that density-split clustering can tighten the constraints on geometry and growth by modelling redshift-space distortions around different environments, compared to the standard two-point clustering analysis. \cite{Paillas2022:2209.04310} expanded on this by quantifying the information content of the full shape of the density-split correlation functions, forecasting that the method can deliver precise constraints on $\Lambda$CDM parameters, and can potentially be used to put upper limits on the sum of neutrino masses.

Until now, a model that can capture the cosmological dependence of the full shape of the density-split correlation functions was not available. In \cite{Cuesta-Lazaro2023:2309.16539}, we have presented \textsc{sunbird}, a simulation-based model for density-split and two-point clustering that can operate down to intra-halo scales, well into the non-linear regime, which has been validated against high-fidelity mock galaxy catalogues based on the \textsc{AbacusSummit} suite of simulations. In this work, we use \textsc{sunbird} to carry out the first application of density-split clustering to observational data, applying it to the final data release of the Baryon Oscillation Spectroscopic Survey \citep{Dawson2013}. We fit a full-shape model of the galaxy two-point correlation function and density-split multipoles down to $1 \Mpch$, putting constraints on the base-$\Lambda$CDM model, as well as on extensions to the base model that vary the dark energy equation of state, the density of relic neutrinos, and the running of the spectral index of the primordial power spectrum.

The paper is organized as follows. We define our observables in Sect.~\ref{sec:observations}. The clustering modelling, as well as the simulations used to validate our pipeline are presented in Sect.~\ref{sec:modelling}. We present our main cosmological constraints and the validation tests in Sect.~\ref{sec:results}. Finally, we summarize and conclude in Sect.~\ref{sec:conclusions}.

\section{Observations}
\label{sec:observations}

\subsection{BOSS CMASS galaxy sample}

We use data from the final data release (DR12) of the Baryon Oscillation Spectroscopic Survey \citep[BOSS, ][]{Dawson2013}. BOSS was a survey conducted as part of the third stage of the larger Sloan Digital Sky Survey \citep[SDSS,][]{York2000:astro-ph/0006396}, which collected optical spectra from more than 1.5 million targets using the 2.5-m Sloan Telescope \citep{Gunn2006:astro-ph/0602326} at Apache Point, New Mexico. BOSS covered roughly 10,000 $\rm{deg}^2$ of the sky in two hemispheres, referred to as the North and the South Galactic caps (NGC and SGC, respectively).

Our analysis is focused on the CMASS galaxy sample, which is dominated by luminous red galaxies (LRG) that were selected on SDSS multicolour photometric observations \citep{Gunn1998:astro-ph/9809085, Gunn2006:astro-ph/0602326}. CMASS is nearly complete down to stellar mass of $M_{*} \approx 10^{11.3}\,{\rm M_{\odot}}$ for $z > 0.45$ \citep{Maraston2013:1207.6114}, and covers a redshift range $0.4 \lesssim z \lesssim 0.7$. For this paper, we impose a more stringent redshift cut $0.45 \leq z \leq 0.6$ to avoid regions where the galaxy number density drops abruptly, which can potentially bias our model predictions. Additionally, we restrict the analysis to the NGC for simplicity. After imposing these restrictions, this results in a sample with a total volume of $\approx 1.4\,(\Gpch)^3$, an \textit{effective} volume of $\approx 1.1\,(\Gpch)^3$, and an average number density of $\approx 3.5 \times 10^{-4} (\hMpc)^3$.

We use the DR12 large-scale structure catalogues provided by the BOSS collaboration\footnote{\href{https://data.sdss.org/sas/dr12/boss/lss/}{https://data.sdss.org/sas/dr12/boss/lss/}.} \citep{Reid2016:1509.06529}. These catalogues include angles and redshifts for each galaxy, which we convert to comoving Cartesian coordinates by adopting a flat-$\Lambda$CDM fiducial cosmology characterized by a matter density parameter $\Omegam = 0.315$, which closely matches the Planck 2018 best-fit cosmology, assuming base-$\Lambda$CDM \citep{Planck2020}. The BOSS collaboration also provides a set of random catalogues that follow the footprint and radial selection of CMASS galaxies, but with no intrinsic clustering, which are used to estimate the overdensity field as described in the following sections.

\subsection{Clustering measurements}

\subsubsection{Two-point clustering}
\label{subsubsec:tpcf}

Galaxy clustering is usually characterized in terms of the two-point correlation function (2PCF), $\xi^{\rm gg}(r)$, which, in its simplest form, quantifies the excess probability ${\rm d}P$ of finding a galaxy in a volume ${\rm d}V$, separated at a distance $r$ from another galaxy, with respect to an unclustered Poisson distribution:
\begin{equation}
    {\rm d}P = n_g \left[1 + \xi^{\rm gg}(r) \right]{\rm d}V \,,
\end{equation}
where $n_g$ is the mean number density of galaxies in the sample. In the presence of redshift-space distortions \citep[RSD,][]{Jackson1972, Kaiser1987} or Alcock-Paczynski distortions \citep[AP,][]{Alcock1979}, the galaxy distribution appears anisotropic to the observer. To capture this anisotropy, a convenient choice is to bin the correlation function in terms of $s$ and $\mu$, where $s$ is the redshift-space pair separation and $\mu$ is the cosine of the angle between the vector connecting the two galaxies and the observer's line of sight.

A number of estimators have been proposed to measure the 2PCF from observational data. A robust and commonly used estimator is the one introduced by \cite{Landy1993},
\begin{equation} \label{eq:2PCF}
    \xi^{\rm gg}(s, \mu) = \frac{\rm GG - 2GR + RR}{\rm RR} \,,
\end{equation}
where $\rm GG$ is the normalized number of galaxy pairs in the $(s, \mu)$ bin, while $\rm GR$ and $\rm RR$ are the normalized galaxy-random and random-random pairs, which make use of the unclustered random catalogues.

It is useful to separate out the different angular components of the ($s, \mu$) correlation function by decomposing it into multipole moments, defined by
\begin{equation}
    \label{eq:multipoles}
    \xi_\ell (s) = \frac{2 \ell + 1}{2} \int_{-1}^1 \rm d \mu \, \xi(s, \mu) P_\ell (\mu),
\end{equation}
with $\rm P_{\ell}$ the Legendre polynomials.

We measure $\xi^{\rm gg}(s, \mu)$ in CMASS using \textsc{pycorr}\footnote{\href{https://github.com/cosmodesi/pycorr}{https://github.com/cosmodesi/pycorr}.}, which is a wrapper around a modified version of the \textsc{CorrFunc} pair-counting code \citep{corrfunc}. We focus the analysis on the monopole ($\xi_0$) and quadrupole ($\xi_2$) moments. The correlation functions are measured in 241 $\mu$ bins from $-1$ to $1$, and radial bins with scale-dependent widths: $1h^{-1}\,{\rm Mpc}$ bins for $s \in  [0, 4]\, h^{-1}{\rm Mpc}$, $3h^{-1}\,{\rm Mpc}$ bins for  $s \in (4, 30]\, h^{-1}{\rm Mpc}$, and
$5h^{-1}\,{\rm Mpc}$ bins for $s \in (30, 151]\, h^{-1}{\rm Mpc}$. Galaxies are appropriately weighted during the pair counting to consider various observational systematics that can bias the clustering measurements.  The total systematic weight for each galaxy is given by
\begin{equation} \label{eq:weights_systot}
    w_{\rm sys, tot} = w_{\rm sys} (w_{\rm fc} + w_{\rm zf} - 1)\, ,
\end{equation}
where $w_{\rm sys}$, $w_{\rm fc}$, and $w_{\rm zf}$ account for imaging systematics, fiber collisions, and redshift failures \citep{Ross2016:1607.03145}. This is additionally multiplied by a weight $w_{\rm FKP}$, which optimally weights the contribution of galaxies based on their redshift-dependent number density \citep{Feldman1994:astro-ph/9304022},
\begin{equation} \label{eq:weights_fkp}
    w_{\rm FKP} = 1/(1 + n(z) P_0) \,,
\end{equation}
where $P_0 = 10^4\, h^3{\rm Mpc}^{-3}$. The total weights for each galaxy are then given by
\begin{equation} \label{eq:weights_tot}
    w_{\rm tot} = w_{\rm FKP} w_{\rm sys, tot}\,,
\end{equation}
while points from the random catalogue are only weighted by $w_{\rm FKP}$.

\begin{figure}
    \centering
    \includegraphics[width=0.9\columnwidth]{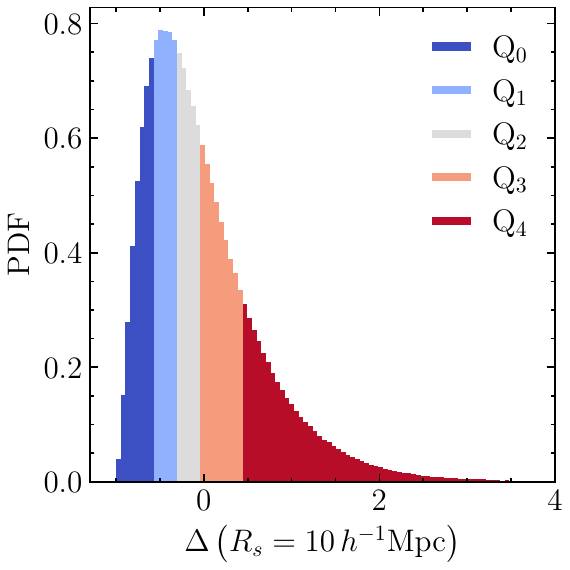}
    \caption{Probability distribution function (PDF) of the galaxy overdensity measured around random query points. The overdensity field has been smoothed with a Gaussian filter of width $R_s = 10\,\Mpch$. The colours represent the split of the PDF into quintiles. \href{https://github.com/florpi/sunbird/blob/main/paper_figures/boss/density_pdf.py}{\faGithub}}
    \label{fig:density_pdf}
\end{figure}

\subsubsection{Density-split clustering}
\label{subsubsec:dsc}

The density-split clustering (DSC) method \citep{Paillas2021} characterizes galaxy clustering in bins of environmental density, with the aim of extracting and combining the cosmological information coming from distinct environments of the cosmic web. We apply the density-split algorithm to the CMASS galaxy sample using our publicly available code\footnote{\href{https://github.com/epaillas/densitysplit}{https://github.com/epaillas/densitysplit}.}, with slight modifications to the algorithm presented in \cite{Paillas2022:2209.04310} to account for the non-uniform survey geometry.

We start by painting the CMASS galaxies and randoms to a rectangular grid that fully encompasses the survey volume, and we estimate the overdensity field as
\begin{equation}
    \delta = \frac{\rm G}{\rm R} - 1\,
\end{equation}
where ${\rm G}$ and ${\rm R}$ are the normalized galaxy and random counts in each cell, weighted as given in Eq.~(\ref{eq:weights_systot}). We smooth the overdensity field with a Gaussian filter of radius $R_s = 10\,h^{-1}{\rm Mpc}$, and then we sample it using cloud-in-cell interpolation at $N_{\rm query}$ query positions, which are taken from the CMASS random catalogue. Here we set $N_{\rm query}$ to 5 times the number of galaxies in the catalogue, and split those query points into 5 quintiles, according to the overdensity at each location.

Figure.~\ref{fig:density_pdf} shows the probability distribution function (PDF) of galaxy overdensity measured at the query positions. The overdensity field shows a non-Gaussian PDF with significant skewness and kurtosis. This shape is a consequence of the growth of structure being bounded at $\Delta = -1$ from below (regions completely devoid of galaxies), while no such constraint is present at the positive $\Delta$ end. The distribution peaks at negative overdensities, reflecting that the average region in the Universe is underdense due to the larger volume occupied by voids. The division into quintiles is demarcated by the different colours in the figure. We label the quintiles as $\Q_i$, where $i$ goes from 0 to 4 from lower to higher densities. In what follows, we discard $\Q_2$ from the clustering analysis, since all five quintiles are not independent from each other.\footnote{As the query points are random, the sum of the density-split cross-correlation functions over quintiles vanishes up to shot noise, and all the information in $\Q_2$ is already contained in the remaining four quintiles, as shown in previous work \citep{Paillas2022:2209.04310}.}

\begin{figure}
    \centering
    \includegraphics[width=0.9\columnwidth]{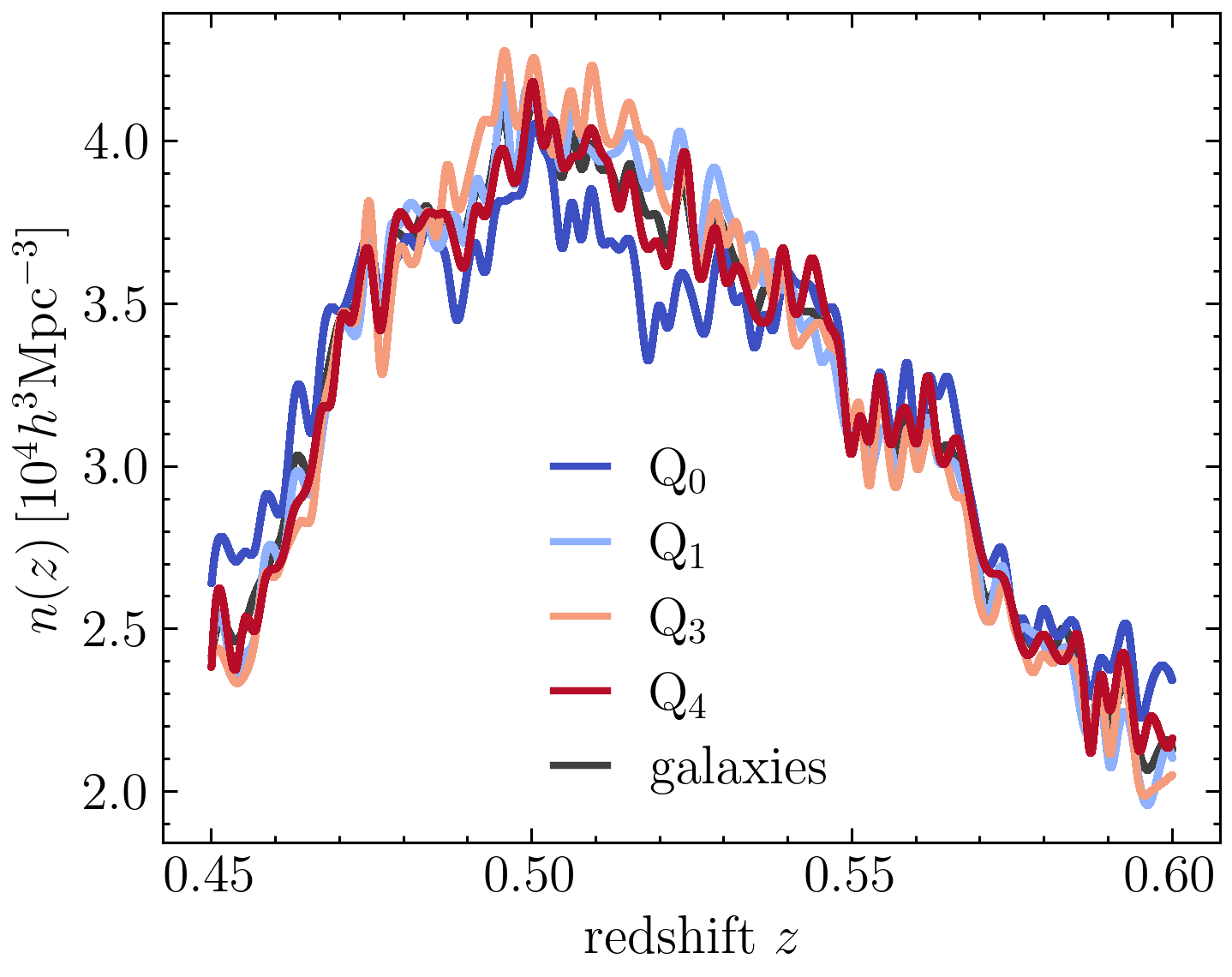}
    \caption{Comoving number density as a function of redshift for the density-split quintiles and galaxies from the BOSS DR12 CMASS sample. \href{https://github.com/florpi/sunbird/blob/main/paper_figures/boss/quantiles_nz.py}{\faGithub}}
    \label{fig:quintiles_nz}
\end{figure}

Figure~\ref{fig:quintiles_nz} shows the comoving number density of quintile positions as a function of redshift, which is observed to closely follow that of the galaxies. This is given by construction, as the query positions are sampled from the clustering random catalogues, which in turn are constructed to match the footprint and radial selection of the galaxies.

Once the density quintiles are defined, we estimate the quintile-galaxy cross-correlation function using the estimator \citep{Landy1993}
\begin{equation} \label{eq:CCF}
    \xi^{\rm qg}(s, \mu) = \frac{\rm QG - QR - RG + RR}{\rm RR} \,,
\end{equation}
where ${\rm QG}$, ${\rm QR}$, ${\rm RG}$, ${\rm RR}$ are the normalized quintile-galaxy, quintile-random, galaxy-random, and random-random pair counts. We note that this assumes that the random catalogue is the same for galaxies and quintiles, which is justified by the good match between the $n(z)$ distributions from Fig.~\ref{fig:quintiles_nz}. The autocorrelation functions of DS quintiles are estimated as 
\begin{equation} \label{eq:ACF}
    \xi^{\rm qq}(s, \mu) = \frac{\rm QQ - 2QR + RR}{\rm RR} \,,
\end{equation}
with $\rm QQ$ being the normalized quintile-quintile pair counts.

We adopt the same binning scheme as for the galaxy 2PCF, as described in Sect.~\ref{subsubsec:tpcf}, and we weight the galaxy pairs according to Eq.~(\ref{eq:weights_tot}).

\begin{figure*}
    \centering
    \begin{tabular}{ccc}
      \includegraphics[width=0.3\textwidth]{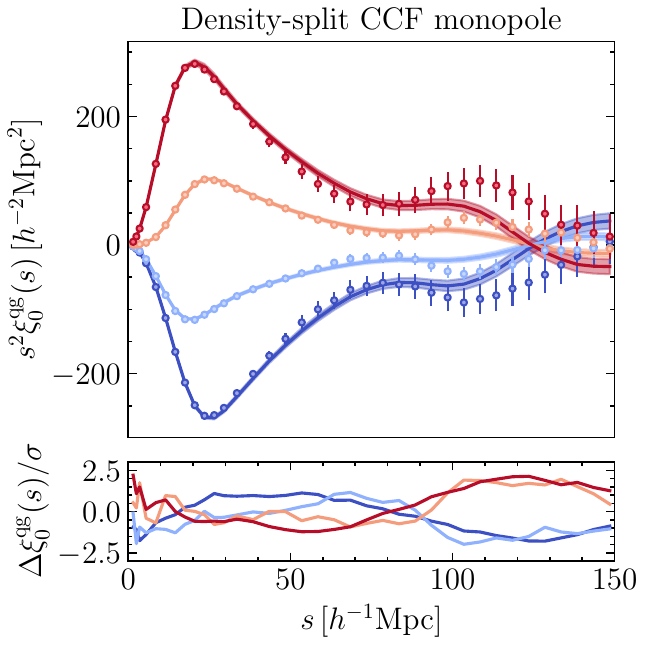}  & \includegraphics[width=0.3\textwidth]
      {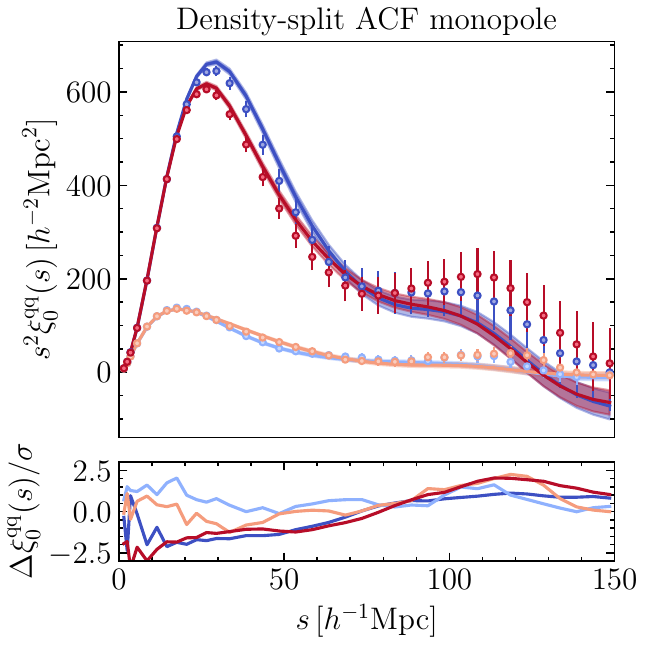} & \includegraphics[width=0.3\textwidth]{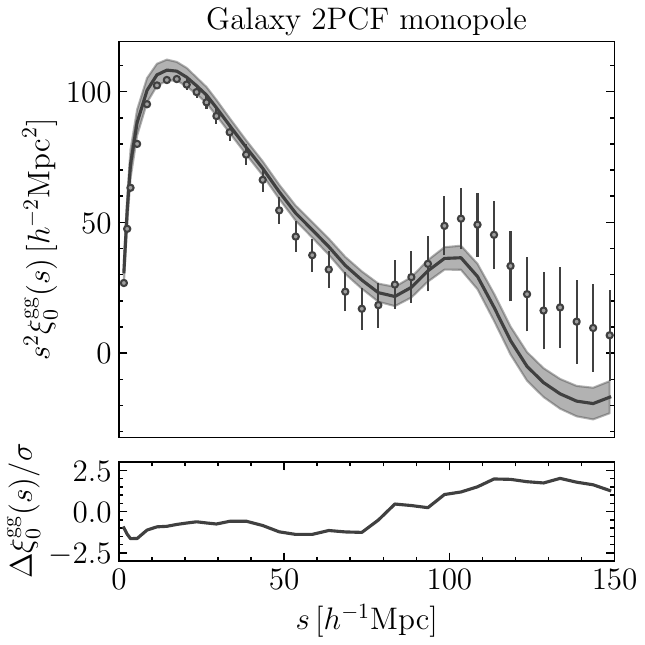}\\[4pt]
      \includegraphics[width=0.3\textwidth]{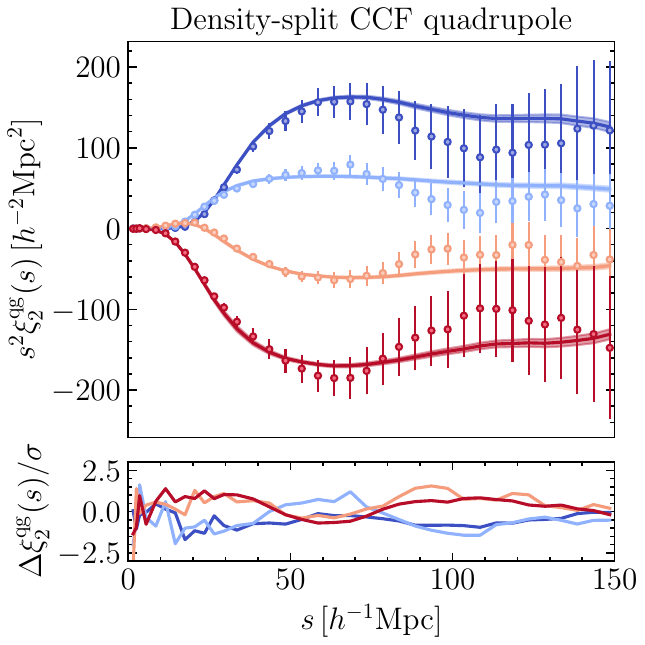}  & \includegraphics[width=0.3\textwidth]
      {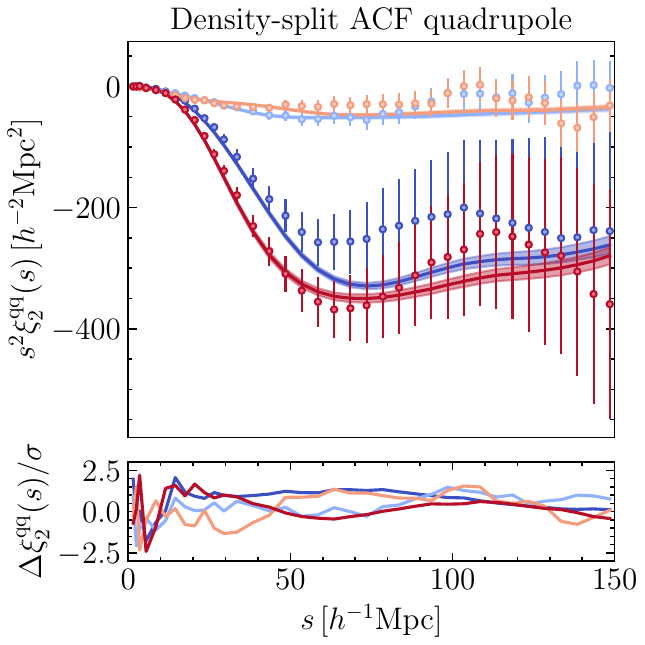} & \includegraphics[width=0.3\textwidth]{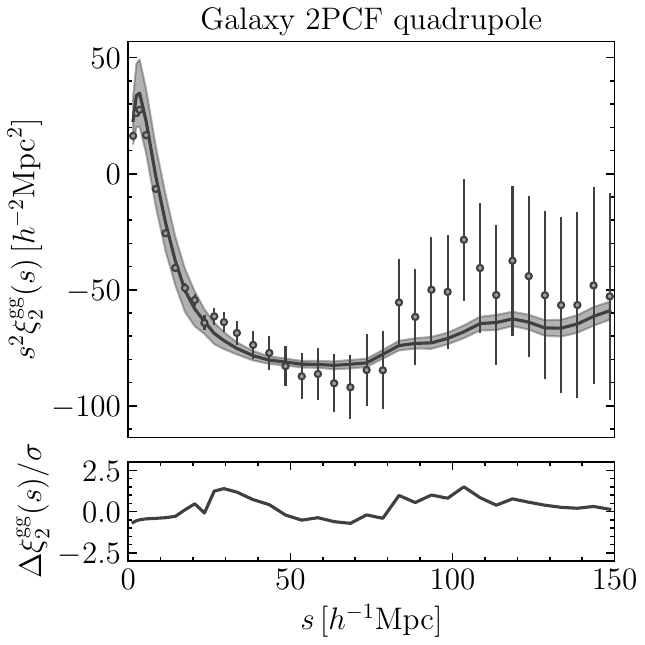}
    \end{tabular}
    \caption{DSC and galaxy 2PCF multipoles measured from the BOSS DR12 CMASS catalogue (circles with error bars), along with the best-fit model from our emulator (solid lines with shaded bands). The columns show the multipoles of the quintile-galaxy cross-correlation function (left), the quintile autocorrelation function (middle), and the galaxy two-point correlation function (right). The lower sub-panels display the difference between the model and the data, in units of the standard deviation of the total error budget. Each colour corresponds to a different density quintile, as illustrated in Fig.~\ref{fig:density_pdf}.\\
    The 68 per cent errors of the data are estimated from 2048 realizations of the MD-Pathchy mocks and represent the expected level of sample variance for the CMASS NGC volume. The emulator uncertainty is estimated by validating the predictions against a test set of simulations with a known cosmology. \href{https://github.com/florpi/sunbird/blob/main/paper_figures/boss/multipoles.py}{\faGithub}}
    \label{fig:multipoles}
\end{figure*}

Figure~\ref{fig:multipoles} shows the measured multipoles from DSC and the galaxy 2PCF. The DSC auto- and cross-correlation functions cover a broad range of amplitudes, as the quantiles trace the underlying matter density field in different ways. $\Q_0$ \& $\Q_1$ are underdense regions with negative linear bias parameters that usually range from -3 to -1, whereas $\Q_3$ \& $\Q_4$ tracer overdensities and have positive bias parameters that can range from 1 to 3 \citep{Paillas2021}. Two signature features are spotted in all profiles: a transition regime around $25 \Mpch$, which is due to the scale that was used to smooth the overdensity field and define the quintiles, and the peak/valley around $100 \Mpch$, which is an imprint of the BAO that originated from sound waves in the photon-baryon plasma prior to recombination.

The quadrupoles show a large degree of anisotropy in DSC. Two main factors contribute to this anisotropy. Firstly, there is an RSD effect caused by the dynamics of galaxies around different density environments. Kaiser-like motions on large scales and random motions on small scales (Fingers of God) cause distinct RSD patterns on the clustering of each quintile, similar to the well-known effects seen in the galaxy 2PCF. Secondly, there is an RSD effect imprinted on the quintile positions themselves, which is the product of identifying the density quintiles in redshift space. \cite{Paillas2022:2209.04310} showed that splitting densities in redshift space causes selection effects that induce distortions in the clustering of the quitiles, which manifests itself as a quadrupole moment in the DSC autocorrelation functions. This is similar in nature to the selection effect that is produced when cosmic voids are identified in redshift space \citep{Chuang2017, Nadathur2019b, Correa2020}.

We reserve the discussion of the model fits (solid lines) for Sect.~\ref{subsec:model_fits}.

\section{Modelling}
\label{sec:modelling}

\subsection{Mock galaxy catalogues}

Throughout this work, we use different types of mock galaxy catalogues to study various aspects of the analysis. The MD-Patchy mocks are used to estimate the sample variance associated to our observed clustering measurements. The Nseries mocks are used to validate the theory model we apply to CMASS. The AbacusSummit mocks are used to train our simulation-based model for galaxy clustering, as presented in our companion paper \citep{Cuesta-Lazaro2023:2309.16539}. Each of these mocks are described in more detail below.

\subsubsection{MD-Patchy mocks}

To estimate the covariance of the data vector, we use the MultiDark-Patchy mocks \citep[MD-Patchy,][]{Kitaura2016:1509.06400}, a suite of 2048 mock galaxy catalogues that were designed to match the footprint, redshift distribution and halo occupation distribution of the BOSS DR12 galaxy samples. The mocks are based on approximate lightcones generated with augmented Lagrangian perturbation theory (ALPT) using the \textsc{patchy} code \citep{Kitaura2014:1307.3285, Kitaura2015:1407.1236}. ALPT is based on a combination of second-order Lagrangian perturbation theory on large scales and the spherical collapse model on smaller scales. \textsc{patchy} populates dark matter haloes using a subhalo abundance matching prescription that is calibrated from N-body simulations from the Big MultiDark Suite \citep{Klypin2016:1411.4001}, and uses a $\Lambda$CDM model matched to the best fit to the Planck 2013 CMB measurements \citep{Planck2014}, characterized by a matter density parameter $\Omega_{\rm m} = 0.307115$, a baryon density parameter $\Omega_{\rm b} = 0.048$, an amplitude of matter fluctuations in $8 \Mpch$ spheres $\sigma_8 = 0.8288$, a tilt of the primordial power spectrum $n_s = 0.9611$, and a dimensionless Hubble parameter $h = 0.6777$.

\subsubsection{Nseries mocks}
\label{subsubsec:nseries}

To study systematic errors in our theory model, we use the Nseries cutsky mocks, a collection of 84 mock catalogues that were designed to match the clustering, footprint and radial selection of the CMASS galaxy sample. The cutsky mocks are constructed from a base set of seven independent full N-body simulations, with a box side length of $2.6 \Gpch$ and a mass resolution of $1.5\times 10^{11} \Msunh$. Dark matter haloes are populated with a halo occupation distribution prescription, with parameters chosen to best model the clustering of the BOSS DR12 CMASS sample. Each box is trimmed and rotated in different ways to produce 12 cutsky mocks that match the geometry of the CMASS sample, which results in a total of 84 pseudo-independent cutsky catalogues. The mocks are then passed through the same fiber assignment code that was used for BOSS, so that they faithfully reproduce the angular variations of fiber collisions in the data \citep{Hahn2017:1609.01714}. 

Nseries is characterized by a cosmology $\Omegam = 0.286$,  $\Omegab = 0.0470$, $h = 0.70$, $\sigma_8 = 0.82$, and $n_s = 0.96$. 

\begin{table*}
    \renewcommand{\arraystretch}{1.2}
    \centering
    \caption{List of cosmological and HOD parameters used in our analysis. For each, we quote the parameter symbol, the prior distribution, the fixed value in the baseline model (where appropriate), and the physical interpretation. We note that the prior distribution for all parameters is uniform, with the exception of the baryon density, for which we adopt a normal distribution with a mean and dispersion as specified.}
    \rowcolors{2}{gray!15}{white}
    \begin{tabular}{| l | l | c | l|}
        \hline
        Parameter & Prior distribution & Baseline & Interpretation\\
        \hline
        $\omega_{\rm b}$ & $\gauss (0.02268, 0.00038)$ & --- &  Physical baryon density\\
        $\omega_{\rm cdm}$ & $\unif [0.1032, 0.14]$ & --- & Physical cold dark matter density\\
        $\sigma_8$ & $\unif [0.687, 0.938]$ & ----& Amplitude of matter fluctuations in $8\,h^{-1}{\rm Mpc}$ spheres\\
        $n_s$ & $\unif [0.901, 1.025]$ & --- & Spectral index of the primordial power spectrum\\
        $\nrun$ & $\unif [-0.038, 0.038]$ & 0.0 & Running of the spectral index\\
        $\Neff$ & $\unif [2.1902, 3.9022]$ & 3.0146 & Number of ultra-relativistic species\\
        $w_0$ & $\unif [-1.27, -0.70]$ & -1.0 & Present-day dark energy equation of state\\
        $w_a$ & $\unif [-0.628, 0.621]$ & 0.0 & Time evolution of the dark energy equation of state\\
        \hline
        $M_{\rm cut}$ & $\unif [12.4, 13.3]$ & --- & Minimum halo mass to host a central \\
        $M_1$ & $\unif [13.2, 14.4]$ & --- & Typical halo mass to host one satellite\\
        $\log \sigma$ & $\unif [-3.0, 0.0]$ & --- & Slope of the transition from hosting zero to one central\\
        $\alpha$ & $\unif [0.7, 1.5]$ & --- & Power-law index for the mass dependence of the number of satellites \\
        $\kappa$ & $\unif [0.0, 1.5]$ & --- & Parameter that modulates the minimum halo mass to host a satellite\\
        $\alpha_c$ & $\unif [0.0, 0.5]$ & --- & Velocity bias for centrals\\
        $\alpha_s$ & $\unif [0.7, 1.3]$ & --- & Velocity bias for satellites\\
        $B_{\rm cen}$ & $\unif [-0.5, 0.5]$ & --- & Environment-based assembly bias for centrals\\
        $B_{\rm sat}$ & $\unif [-1.0, 1.0]$ & --- & Environment-based assembly bias for satellites \\
        \hline
    \end{tabular}
    \label{tab:priors}
\end{table*}

\subsubsection{AbacusSummit mocks}

To train and validate our clustering emulators, we use \textsc{AbacusSummit}, a suite of cosmological N-body simulations \citep{Maksimova2021:2110.11398} designed to meet the simulation requirements of the Dark Energy Spectroscopic Instrument \citep{Levi2019}. The simulations were run with the \textsc{Abacus} N-body code \citep{Garrison2019:1810.02916, Garrison2021:2110.11392}, comprising different volumes, resolutions, and cosmologies. The \textit{base} simulations follow the evolution of $6912^3$ dark matter particles in a $(2\Gpch)^3$ volume, with a mass resolution of $2 \times 10^9 \Msunh$. There are 97 cosmology variations in total, exploring an eight-dimensional parameter space around the Planck18 primary $\Lambda$CDM cosmology \citep[PL18,][]{Planck2020}:
\begin{align} \label{eq:abacus_parameters}
    \bm{\theta}_{\rm \textsc{AbacusSummit}} = \{ \omegac, \omegab, \sigma_8, n_s, \nrun, \Neff, w_0, w_a \} \,.
\end{align}
Here, $\omegac = \Omega_{\rm c}h^2$ and $\omegab = \Omega_{\rm b}h^2$ are the physical cold dark matter and baryon densities, $\nrun$ is the running of the spectral tilt, $\Neff$ is the effective number of ultra-relativistic species, $w_0$ is the present-day dark energy equation of state, and $w_a$ captures the time evolution of the dark energy equation of state. The simulations assume a flat spatial curvature, and the dimensionless Hubble parameter $h$ is calibrated to match the Cosmic Microwave Background (CMB) acoustic scale $\theta_*$ to the PL18 measurement.

For the training and validation of the emulators, we restrict to the following subset of simulations, which were all run with the same initial conditions:
\begin{itemize}
    \item[] \fbox{\texttt{c000}} PL18 $\Lambda$CDM base cosmology, matching the mean of the base\_plikHM\_TTTEEE\_lowl\_lowE\_lensing likelihood.

    \item[] \fbox{\texttt{c001-004}} Secondary cosmologies, including a low $\omega_{\rm cdm}$ choice \citep[WMAP7,][]{WMAP7}, a $w$CDM choice, a high $\Neff$ choice, and a low $\sigma_8$ choice.

    \item[] \fbox{\texttt{c013}} Cosmology matching the Euclid Flagship2 $\Lambda$CDM simulation (Castander et al., in preparation).
    
    \item[] \fbox{\texttt{c100-126}} Linear derivative grid providing paired simulations with small negative and positive steps in the eight-dimensional cosmological parameter space from Eq.~(\ref{eq:abacus_parameters}).

    \item[] \fbox{\texttt{c130-181}} An emulator grid around the \texttt{c000} cosmology that provides a wider coverage of the cosmological parameter space.
\end{itemize}

In addition to the base simulations, there are multiple realizations of smaller boxes with a side length of $500 \Mpch$ at the \texttt{c000} cosmology, which can be used for covariance estimation. Throughout the rest of the paper, we will refer to these simulations as \textsc{AbacusSmall}, and to the \textit{base} simulations simply as \textsc{AbacusSummit}.

Group finding is done on the fly, using a hybrid Friends-of-Friends/Spherical Overdensity algorithm, dubbed CompaSO \citep{Hadzhiyska2021:2110.11408}. As described in Sect.~\ref{sec:emulator}, we populate these halo catalogues with galaxies using a halo occupation distribution prescription, which are then used to obtain the clustering measurements for our training data.

\subsection{Galaxy-halo connection model}
\label{subsec:hod}

In the current paradigm of cosmology, galaxies are thought to form and evolve within dark matter halos, which are large structures that form as a result of the gravitational collapse of overdensities in the Universe. The halo occupation distribution (HOD) is a statistical model that describes how galaxies are distributed within dark matter halos. 

A well-suited model for LRG is the base halo model from \cite{Zheng2007}, where the average number of central galaxies in a halo of mass $M$ is given by
\begin{align}
    \langle N_{\rm c} \rangle(M) = \frac{1}{2} \left(1 + \mathrm{erf} \left(\frac{\log M - \log M_{\rm cut}}{\sqrt{2} \sigma} \right)  \right)\,
\end{align}
where $\mathrm{erf}(x)$ denotes the error function, $M_{\rm cut}$ is the minimum mass required to host a central, and $\sigma$ is the slope of the transition between having zero to one central galaxies. The average number of satellite galaxies is in turn given by
\begin{align}
    \langle N_{\rm s} \rangle(M) = \langle N_{\rm c} \rangle(M) \left(\frac{M - \kappa M_{\rm cut}}{M_1} \right)^{\alpha}\,
\end{align}
where $\kappa M_{\rm cut}$ gives the minimum mass required to host a satellite, $M_1$ is the typical mass that hosts one satellite, and $\alpha$ is the power law index for the number of galaxies.

We use the \textsc{AbacusHOD} package, which is highly efficient and contains a wide range of HOD extensions \citep{Yuan2022:2110.11412}. For this work, we extend the base model to modulate galaxy peculiar velocities via the parameters $\alpha_{\rm vel, c}$, which parametrizes the velocity bias between the central galaxy and the halo centre, and $\alpha_{\rm vel, s}$, which parametrizes the velocity bias between the satellite galaxies and the local dark matter particles. When no velocity bias is present, $\alpha_{\rm vel, c} = 0$ and $\alpha_{\rm vel, s} = 1$, in which case centrals perfectly follow the velocity of halo centres, and satellites perfectly match the velocity of the underlying dark matter particles.

We introduce two additional parameters to account for galaxy assembly bias: $B_{\rm cen}$ and $B_{\rm sat}$, which add environment-based secondary bias for centrals and satellites, respectively. Here, the environment is defined as the smoothed matter density around the halo centres, using a top-hat filter of radius $R_s = 5\,h^{-1}{\rm Mpc}$. When no secondary bias is present, $B_{\rm cen} = B_{\rm sat} = 0$. Positive/negative values of these parameters indicate a preference for galaxies to form in haloes around less/more dense environments, respectively.

\subsection{Clustering emulators}
\label{sec:emulator}

\begin{figure*}
    \centering
    \includegraphics[width=0.8\textwidth]{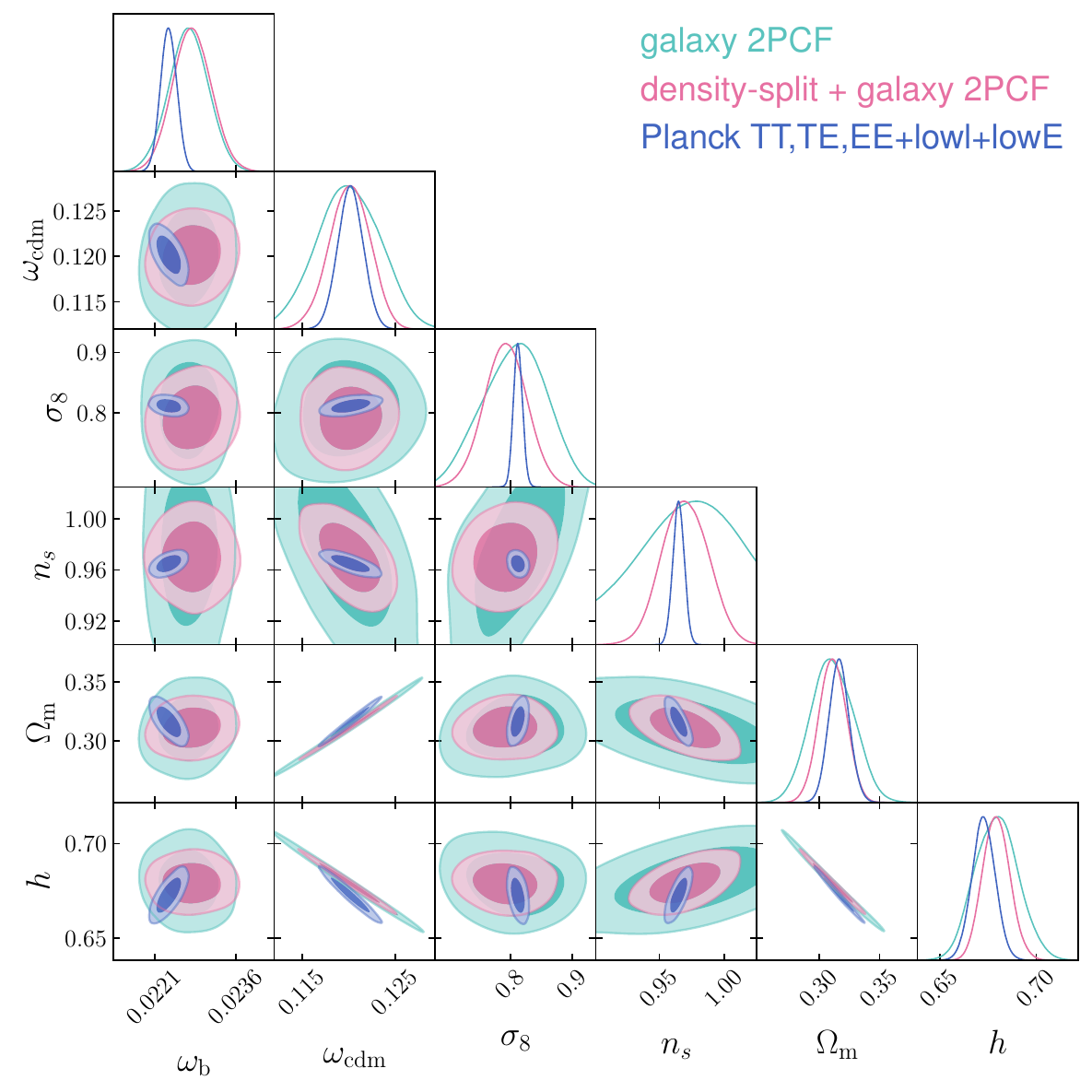}
    \caption{Posterior probability distributions of the base-$\Lambda$CDM parameters from the fits to the BOSS DR12 CMASS clustering data. The pink contours show results from the combination of density-split clustering and the galaxy 2PCF, while the aquamarine contours show results using only the latter. We also overplot constraints taken from the \textsc{Planck\_TTTEEE\_lowl\_lowE} likelihood \citep[][]{Planck2020} in blue. All 2D contours show 68 and 95 per cent confidence intervals. \href{https://github.com/florpi/sunbird/blob/main/paper_figures/boss/cosmo_inference_cmass.py}{\faGithub}}
    \label{fig:cosmo_inference_baseline}
\end{figure*}

In the context of galaxy clustering and cosmology, an emulator refers to a computational model or algorithm that approximates the predictions of expensive or time-consuming simulations or calculations. They are used to efficiently and accurately generate predictions from cosmological models without  resorting to additional simulations.

Emulators are trained on a set of pre-computed simulations, often referred to as a training or calibration set, which cover a wide range of model parameter values and capture the desired properties of interest of an observable. As such, the emulator can learn how this observable responds to changes in cosmology or galaxy-halo connection parameters. Once the emulator is trained, it can rapidly predict the desired model predictions for any given set of input parameters.

In \cite{Cuesta-Lazaro2023:2309.16539}, we present our galaxy 2PCF and DSC emulators, which are based on HOD catalogues constructed from the \textsc{AbacusSummit} simulations. Here, we present a brief description of how the emulators are constructed.

We start from the dark matter halo catalogues from the \textsc{AbacusSummit} snapshots at $z = 0.5$, spanning 85 different cosmologies within the eight-dimensional $w_0w_a\Lambda$CDM parameter space defined in Eq.~(\ref{eq:abacus_parameters}). Using the \textsc{abacushod} code \citep{Yuan2022:2110.11412}, we populate dark matter haloes with a nine-parameter extended HOD framework (Sect.~\ref{subsec:hod}),
\begin{align}
    \bm{\theta}_{\rm HOD} = \{ M_{\rm cut}, M_1, \sigma, \alpha, \kappa, \alpha_{\rm vel, c}, \alpha_{\rm vel, s}, B_{\rm cen}, B_{\rm sat} \}\, ,
\end{align}
generating 100 unique HOD variations per cosmology, where the HOD parameters are sampled from a Latin Hypercube to ensure an optimal sampling of the parameter space. When the number density of an HOD catalogue is above the average number density of the CMASS sample $(n_{\rm gal} \approx 3.5\times 10^{-4}\,(h/{\rm Mpc})^{3})$, we invoke an incompleteness parameter $f_{\rm ic}$ to downsample the catalogue down to the target number density.

Under the distant-observer approximation, we map the positions of galaxies to redshift space by perturbing their positions with their peculiar velocities along one cartesian axis of the simulation box chosen to represent the line of sight. We repeat this procedure for each of the three axes of the simulation boxes, effectively generating three pseudo-independent redshift-space catalogues for each HOD variation, from which we can average the clustering measurements to reduce cosmic variance later on.

For each of these mock catalogues, we run the DSC pipeline, using a mesh resolution of $R_{\rm cell} = 5\,h^{-1}{\rm Mpc}$, a number of random query points $N_{\rm query}$ equal to five times the number of galaxies in each catalogue, a smoothing radius for the Gaussian filter of $R_s = 10\,h^{-1}{\rm Mpc}$, and five density quintiles. We compute the galaxy 2PCF and the density-split correlation functions in bins of $s$ and $\mu$, and we decompose them into their multipole moments.

We split the HOD catalogues into training, validation and test sets, and for each clustering statistic we construct separate fully-connected neural networks, which take the cosmological and HOD parameters as an input, and return the monopole and quadrupole moments of the correlation functions. For training and validation, we use cosmologies \fbox{\texttt{c100-126}} and \fbox{\texttt{c130-181}}, whereas the rest are reserved for the test set. The hyperparameters of the neural networks are calibrated to minimize the validation loss. Overall, we observe that the emulators produce model predictions with percent-level accuracy for the full range of scales, when tested against the test simulation boxes.

\subsection{Likelihood}

When fitting the emulator to the CMASS data, we define the log-likelihood as

\begin{equation}
    \mathcal{L} = \left(\bm{d}^{\rm data} - \bm{d}^{\rm model} \right){\bf C}^{-1} \left(\bm{d}^{\rm data} - \bm{d}^{\rm model} \right)^\top \,
\end{equation}
where $\bm{d}^{\rm data}$ is the observed data vector, $\bm{d}^{\rm model}$ is the emulator prediction, and ${\bf C}$ is the covariance matrix, which includes three contributions to the error budget:
\begin{equation} \label{eq:total_covariance}
    {\bf C} = {\bf C}^{\rm data} + {\bf C}^{\rm emu} + {\bf C}^{\rm sim} \, .
\end{equation}
Here, ${\bf C}^{\rm data}$ is the term associated with the sample variance of the data vector, which is estimated from multiple realizations of the MD-Patchy mocks:
\begin{align} \label{eq:cov_data}
    {\bf C}^{\rm data} = \frac{1}{N_{\rm patchy} - 1} \sum_{k=1}^{N_{\rm patchy}} \left({\bm{d}_{k}} - \overline{\bm{d}}\right)\left({\bm{d}_{k}} - \overline{\bm{d}}\right)^\top \, ,
\end{align}
where $N_{\rm patchy} = 2048$.

The simulations used for training are at a fixed phase, which could be different from the true underlying phase of the Universe. In other words, the cosmic variance is frozen in our emulator predictions. To account for this, we add an extra sample variance contribution to the error budget, associated with the finite size of the training simulations. We estimate this covariance using multiple mock realizations of the AbacusSummit fiducial cosmology with a fixed set of HOD parameters with high likelihood:
\begin{align} \label{eq:cov_sim}
    {\bf C}^{\rm sim} = \frac{1}{N_{\rm sim} - 1} \sum_{k=1}^{N_{\rm sim}} \left({\bm{d}_{k}} - \overline{\bm{d}}\right)\left({\bm{d}_{k}} - \overline{\bm{d}}\right)^\top \, ,
\end{align}
where $N_{\rm sim} = 1800$.

${\bf C}^{\rm emu}$ accounts for the intrinsic error in the model predictions due to an imperfect emulation. This term is calculated by computing a covariance matrix from the difference between the emulator predictions and measurements from a set of test simulations with known cosmologies and HOD parameters, $\Delta \bm{d}$:
\begin{equation} \label{eq:cov_emu}
    {\bf C}^{\rm emu} = \frac{1}{N_{\rm test} - 1}\sum_{k=1}^{N_{\rm test}}\left( \Delta \bm{d}_{k} - \overline{\Delta\bm{d}_{k}} \right) \left( \Delta \bm{d}_{k} - \overline{\Delta \bm{d}_{k}} \right)^\top\, ,
\end{equation}
where the overline denotes the mean across all test simulations, and $N_{\rm test} = 600$.

When calculating the likelihood, both ${\bf C}^{\rm data}$ and ${\bf C}^{\rm sim}$ are multiplied by a factor $\mathcal{P}$ before inversion \citep{Percival2022}:
\begin{align}
    \mathcal{P} = \frac{(N_s - 1)[1 + B(N_d - N_\theta)]}{N_s - N_d + N_\theta - 1} \,,
\end{align}
where
\begin{align}
    B = \frac{N_s - N_d - 2}{(N_s - N_d - 1)(N_s - N_d - 4)} \,.
\end{align}
Here, $N_s$ is the number of simulations used to estimate the covariance, $N_d$ is the length of the data vector and $N_\theta$ is the number of parameters that are being fitted. This corrects for the fact that even though Eq.~(\ref{eq:cov_data}) \& (\ref{eq:cov_sim}) are unbiased estimates of the covariance matrices, the inversion leads to biased parameter constraints.

We sample the posterior distribution of parameters using the \textsc{dynesty} nested sampler \citep{dynesty:1904.02180}, which also provides an estimation of the Bayesian evidence. As listed in Table~\ref{tab:priors}, we assume flat priors for all parameters except for the baryon density, for which we adopt a BBN-informed Gaussian prior \citep{Aver2015:1503.08146, Cooke2018:1710.11129, Schoneberg2019:1907.11594} for our baseline analysis:
\begin{equation}
    \omega_b = 0.02268 \pm 0.00038\, .
\end{equation}
In later sections, we also explore using a flat prior range for $\omega_{\rm b}$, finding that it has little impact on our conclusions and the reported constraints for other parameters.

\begin{table}
    \renewcommand{\arraystretch}{1.4}
    \centering
    \caption{The $\chi^2$, degrees of freedom, log-likelihood, and log-evidence for the best fits to the BOSS DR12 multipoles from the galaxy two-point correlation function (2PCF), density-split clustering (DSC) and the combination of the two. We simultaneously fit the monopole and quadrupole, using scales $1 \Mpch < s < 150 \Mpch$.}
    \rowcolors{2}{white}{gray!15}
    \begin{tabular} {|c|c|c|c|c|c|}
    \hline
    Statistic & $\chi^2$ & ${\rm dof}$ & $\chi^2/{\rm dof}$ & $\log \mathcal{L}$ & $\log \mathcal{Z}$\\
    \hline
    2PCF + DSC & 1052 & 635 & 1.66 & -257.58 & -276.38\\
    DSC        & 942 & 563 & 1.67 & -253.15 & -270.85\\
    2PCF       & 63 & 59 & 1.06 & -29.63 & -38.03 \\
    \hline
    \end{tabular}
    \label{tab:chi2}
\end{table}

\subsection{$\theta_*$ prior}
\label{sec:theta_*}

In the simulations used to train our emulator, the value of the dimensionless Hubble parameter $h$ was chosen to fix the CMB acoustic scale $\theta_*$ ($100\theta_*=1.041533$) matching the best-fit measurement from PL18 \citep{Planck2020,Maksimova2021:2110.11398}. This means that effectively we apply a fixed $\theta_*$ constraint as a prior on our emulator results. This ensures that we only consider models within the part of parameter space where the emulator has been trained.

As a result, $h$ is not a free parameter in our model, but is determined by the sampled parameters and the $\theta_*$ constraint. We use \textsc{class} \citep{class:1104.2933} to derive $h$ at each point in our chains, and we then use this to also obtain $\Omegam$, calculated as
\begin{equation}
    \Omegam = (\omegab + \omegac + \omega_{\nu})/h^2\,,
\end{equation}
where $\omega_{\nu} = 0.00064420$ accounts for 60 meV neutrinos (also the choice of the base PL18 cosmology), which is always fixed in our model.

\begin{table*}
    \renewcommand{\arraystretch}{1.4}
    \centering
    \caption{Parameter constraints for the base $\Lambda$CDM fits on the BOSS DR12 CMASS galaxy catalogue, using the combination of density-split clustering and the galaxy 2PCF, or using only the latter.}
    \rowcolors{2}{gray!15}{white}
    \begin{tabular} {|c |c c |c c |c c|}
    \hline
    & \multicolumn{2}{c|}{density-split + galaxy 2PCF} & \multicolumn{2}{c|}{density-split} & \multicolumn{2}{c|}{galaxy 2PCF} \\
     \hline
     Parameter &  best-fit & mean $\pm \sigma$ & best-fit & mean $\pm \sigma$ & best-fit & mean $\pm \sigma$\\
    \hline
    {$\omega_{\rm b} $}  & $0.0234$ & $0.02279\pm 0.00035$ & $0.0233$  & $0.02283\pm 0.00035$         & $0.0226$  & $0.02272\pm 0.00036$\\
    {$\omega_{\rm cdm}$} & $0.1209$ & $0.1201\pm 0.0022$   & $0.1187$  & $0.1191^{+0.0026}_{-0.0022}$ & $0.1190$  & $0.1200\pm 0.0034$\\
    {$\sigma_8       $}  & $0.8116$ & $0.792\pm 0.034$     & $0.7715$  & $0.768^{+0.037}_{-0.043}$             & $0.7591$  & $0.807^{+0.055}_{-0.049}$\\
    {$n_s            $}  & $0.9837$ & $0.970\pm 0.018$     & $0.9952$  & $0.968\pm 0.023$    & $0.9515$  & $0.969^{+0.043}_{-0.026}$\\
    \hline
    $h$                & $0.6774$  & $0.6793\pm 0.0070$    & $0.6846$ & $0.6828^{+0.0070}_{-0.0083}$     & $0.6832$ & $0.679\pm 0.011$\\
    $\Omegam$            & $0.3157$   & $0.311\pm 0.011$   & $0.3044$  & $0.306\pm 0.012$            & $0.3047$  & $0.311\pm 0.018$\\
    $f\sigma_8$          & $0.4747$   & $0.462\pm 0.020$   & $0.4482$  & $0.447^{+0.021}_{-0.025}$   & $0.4411$  & $0.470^{+0.033}_{-0.028}$\\
    \hline
    \end{tabular}
    \label{tab:base_lcdm_constraints}
\end{table*}

A consequence of this prior on $\theta_*$ is that the parameter constraints we obtain below do not come exclusively from late-Universe clustering measurements but rather from a combination of galaxy clustering and information on the CMB acoustic scale, although they do not use other CMB information. The simple geometrical interpretation of $\theta_*$ means that it is one of the best-measured quantities in all cosmology (the PL18 measurement corresponding to a 0.03 per cent precision level), and this measurement is also extremely robust to changes in the cosmological model \citep{Planck2020}. We therefore consider this to be a very well-justified prior.

\section{Results}
\label{sec:results}

In this section, we apply our emulator framework to the CMASS clustering data and present our main cosmological constraints, along with tests for model systematics. We focus on the constraints for cosmological parameters and reserve the discussion about HOD constraints for Appendix~\ref{ap:hod_constraints}.

\subsection{Model fits}
\label{subsec:model_fits}

The solid lines in Fig.~\ref{fig:multipoles} show the best-fit base-$\Lambda$CDM model to the measured multipoles, with the $\chi^2$, likelihood, and Bayesian evidence values reported in Table~\ref{tab:chi2}. The $\chi^2$ per degree of freedom for the galaxy 2PCF + DSC combination is 1.66, while for the 2PCF-only fit it is 1.06, which is comparable to the reported values for the small-scale 2PCF fit from \cite{Yuan2022b:2203.11963} and the configuration-space BAO fit from \cite{Ross2016:1607.03145}.

We observe some hints of the models under-predicting (or over-predicting, in the case of negative density contrasts) the observed clustering at scales $s > 90 \Mpch$. Such excess large-scale clustering has been observed before in the galaxy 2PCF analysis of BOSS data.  In \cite{Ross2016:1607.03145}, the authors find that the observed monopole shows an apparent excess with respect to the mean of the MD-Patchy mock catalogues, although it is argued that the mismatch is of low statistical significance. In \cite{Satpathy2017:1607.03148}, the authors fit the correlation function using a model based on Convolutional Lagrangian Perturbation Theory, also finding that the model underpredicts the monopole data at large scales. Here, we observe a similar trend when comparing the measured multipoles to our emulator predictions, although the level of discrepancy is never greater than two standard deviations for all scales considered, irrespective of the summary statistic. As informed by the $\chi^2$ values, the quality of the fit is better than might be guessed by the eye, since the separation bins are correlated. In Sect.~\ref{subsec:systematics}, we show that the mean clustering signal of the Nseries mocks monopole is also lower than the CMASS data, although still consistent within $2\sigma$. We also observe that the emulator is able to fit the mean of the mocks with much higher accuracy than the data, which we deem reasonable given that our model was trained to make predictions for the ensemble average of the clustering statistics.

Although some of the observed offsets between our best-fit model and CMASS could be partially attributed to statistical fluctuations, we should bear in mind the possibility that there are residual systematic effects in the clustering data that are not fully corrected by the weighting procedure. \cite{Lavaux2019:1909.06396} performed a field-level inference of BOSS data, finding evidence of residual systematics on the spectroscopic data that so far remain unexplained, which can induce correlated modulations of the order of 30 per cent on the sky for CMASS. In addition, it is worth noting that the systematic weights for the large-scale structure catalogues provided by the BOSS collaboration were mainly validated for two-point statistics, and it is possible that they are sub-optimal for alternative clustering methods such as DSC. We plan to explore this topic further in future work.

\subsection{Base-$\Lambda$CDM constraints}
\label{subsec:base_lcdm_constraints}

In this section, we explore the constraining power of DSC and the galaxy 2PCF on the base $\Lambda$CDM parameters, letting $\omegab$, $\omegac$, $\sigma_8$, $n_s$, and the HOD parameters vary during the fit. To generate the model predictions, we query the emulator fixing the remaining cosmological parameters to their baseline values $w_0 = -1$, $w_a = 0$, $\nrun = 0$, and $\Neff = 3.0146$.

Figure~\ref{fig:cosmo_inference_baseline} shows the 2D posterior distributions on $\Lambda$CDM, marginalized over the HOD parameters. We also over-plot the base-$\Lambda$CDM constraints from the \textsc{Planck\_TTTEEE\_lowl\_lowE} likelihood from PL18 to facilitate comparison. The reported best-fit values, along with the means of the marginalized posteriors and their dispersion are listed in Table~\ref{tab:base_lcdm_constraints}. It is worth noting that the best-fit values can be shifted with respect of the mean of the posterior due to its non-Gaussian shape.

The galaxy 2PCF by itself constrains $\omegac$, $\sigma_8$, and $n_s$ with a 2.7, 6.2, and 3.2 per cent precision, respectively. Adding the DSC multipoles tightens the constraining power to a precision of 1.7, 3.8 and 1.8, respectively. The constraints for baryon density are entirely dominated by the BBN prior. Overall, both posterior distributions are largely consistent with the PL18 results, with a $0.04\sigma$ and $0.6 \sigma$ difference in the mean values of $\omegac$ and $\sigma_8$ between our baseline fit and PL18.

As described in Sect.~\ref{sec:theta_*}, we use the $\theta_*$ constraint to obtain $h$ and $\Omegam$ as derived parameters from our chains, and also show the results for them in Fig.~\ref{fig:cosmo_inference_baseline}. We obtain $h = 0.6793\pm 0.0070$ and $\Omegam = 0.3122^{+0.0094}_{-0.011}$ when fixing to the base $\Lambda$CDM model. As Fig.~\ref{fig:cosmo_inference_baseline} shows, we do not observe any significant degeneracy between $\Omegam$ and the baryon density (and our constraints on $\omegab$ are also prior-dominated). This means that, to a good approximation, $\theta_*\propto\Omegam h^{3.4}$, the expected behaviour for fixed $\Omegab h^2$ within flat-$\Lambda$CDM models \citep{Percival2002:astro-ph/0206256}, and thus DSC and the galaxy 2PCF results both describe this narrow degeneracy direction in the $\Omegam$-$h$ plane. This is close to but not the same as the similarly tight degeneracy obtained from the the CMB by PL18, which corresponds to constant $\Omegam h^3$. The degeneracy direction for models with the same $\theta_*$ depends weakly on $\Omegab h^2$ through the sound horizon. PL18 uses the CMB data to also constrain $\Omegab h^2$, leading to an anti-correlation between constraints on $\Omegab$ and $\Omegam$, which then results in a different degeneracy in the $\Omegam$-$h$ plane.

\begin{figure*}
    \centering
    \includegraphics[width=0.85\textwidth]{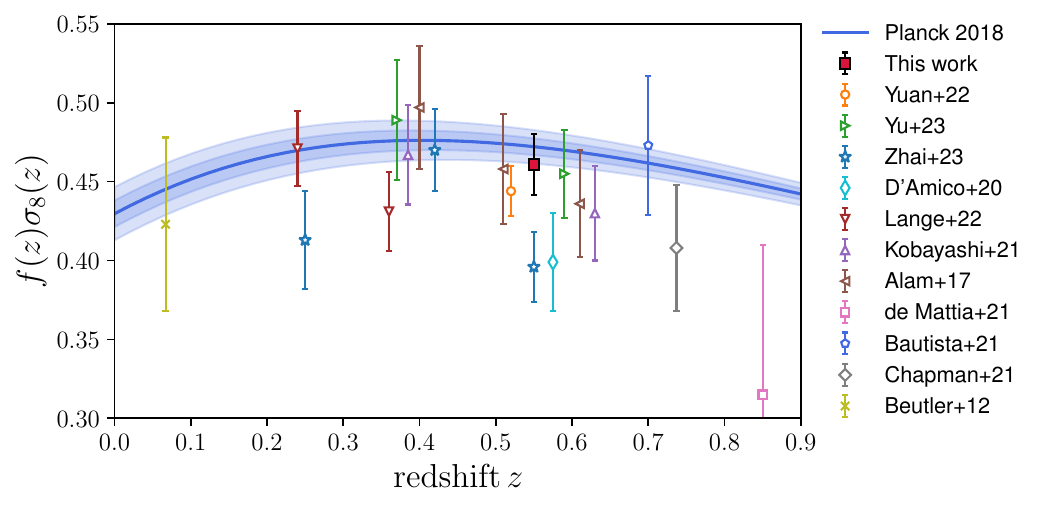}
    \caption{Growth rate of structure derived from our analysis of the BOSS DR12 CMASS sample, combining density-split clustering and galaxy 2PCF measurements. The base-$\Lambda$CDM prediction, based on the best-fit Planck 2018 cosmology, is shown in blue, where the darker and lighter shades represent 68 and 95 per cent confidence intervals. We also compare with other measurements from the literature at different redshifts. The data include 6dFGS \citep{Beutler2012:1204.4725}, eBOSS \citep{deMattia2021:2007.09008, Bautista2021, Chapman2022:2106.14961}, as well as other studies performed on BOSS: the consensus DR12 analysis \citep{Alam2017:1607.03155}, a LOWZ simulation-based analysis \citep{Lange2022:2101.12261}, CMASS simulation-based analyses \citep{Yuan2022b:2203.11963, Kobayashi2022:2110.06969, Zhai2023:2203.08999}, an EFTofLSS analysis \citep{d'Amico2020:1909.05271}, and a halo perturbation theory analysis \citep{Yu2023:2211.16794}. In some cases, the redshifts of the measurements have been slightly shifted horizontally for visual clarity. \href{https://github.com/florpi/sunbird/blob/main/paper_figures/boss/fs8_cmass.py}{\faGithub}}
    \label{fig:fs8}
\end{figure*}

Within $\Lambda$CDM, the linear growth rate of structure can be approximated as
\begin{align}
    f(z) \approx \Omegam^{0.55}(z) \,.
\end{align}
We use this expression to derive a value for $f\sigma_8$ at the effective redshift of CMASS from our chains\footnote{As we are doing a direct fit of $\Lambda$CDM parameters and marginalizing over $h$, our results are not affected by the problems of the standard template-based clustering analyses pointed out in \cite{Sanchez2020}.}. We obtain $f\sigma_8(z=0.525) = 0.462\pm 0.020$, which is a 4.3 per cent constraint from the combination of DSC and the galaxy 2PCF. We contrast this constraint with the PL18 prediction and other clustering studies in Fig.~\ref{fig:fs8}. Our constraint is slightly lower than the mean of PL18, but consistent at the $1\sigma$ level. We also find good agreement with the results of the main BOSS clustering analysis \citep{Alam2017:1607.03155}, but we get a factor of 1.9 better precision using the DSC + galaxy 2PCF combination, compared to their middle redshift bin that covers $0.4 < z < 0.6$.

We find a higher mean $f\sigma_8$ than \citet{Yuan2022b:2203.11963} (from hereon Y22), although there is still consistency at the 0.7$\sigma$ level. Interestingly, Y22 constrain $f\sigma_8$ to a precision similar to our baseline analysis, even though they only use the galaxy 2PCF at scales $< 30 \Mpch$. Our 2PCF-only fit is a factor of 1.8 worse than theirs, which seems puzzling given that their model is based on the same set of simulations used in this study. The main reason we attribute this difference to is that while we estimate the intrinsic emulator error using a test set of simulations that covers the whole prior range in the cosmology \& HOD parameter space, Y22 estimate the error using only HOD catalogues that have a high likelihood with respect to the measured data vector. Our emulator error will generally tend to be more conservative, as it also considers the error around regions near the edge of the priors, which are usually harder to emulate. It also dominates the error budget at small scales, precisely where Y22 extract most of their information. Effectively, this means we are discarding some small-scale information which they keep.  We have verified that by artificially lowering or removing the emulator error, the precision of our galaxy 2PCF constraints closely matches that of Y22. Other reasons that could potentially add to this are that Y22 use Gaussian priors on the HOD parameters while we use flat priors, and that Y22 estimate the data covariance through jackknife resampling, while our covariance is estimated from mock catalogues.

Our results are $2.2\sigma$ higher than those reported by \cite{Zhai2023:2203.08999}, who find $f\sigma_8(z = 0.55) = 0.396 \pm 0.022$. \cite{Zhai2023:2203.08999} train an emulator using the \textsc{Aemulus} suite of simulations to fit the small-scale clustering of BOSS galaxies using the galaxy 2PCF between $0.1 \Mpch$ and $60 \Mpch$. Although we have not tested our emulator against \textsc{Aemulus} directly, our companion paper \citep{Cuesta-Lazaro2023:2309.16539} shows that our model can recover unbiased cosmological constraints when fitted to mock galaxy catalogues generated with a different N-body code and galaxy-halo connection model, specifically galaxies generated with the subhalo abundance matching technique by \cite{Zhai2023:2203.08999} from the Uchuu suite of simulations \citep{Ishiyama2021:2007.14720, Dong-Paez2022:2208.00540, Oogi2023:2207.14689, Aung2023:2209.12918, Prada2023:2304.11911}. Applying our emulator to \textsc{Aemulus} could be a useful step to explore in future work to better understand the source of this discrepancy.

\cite{Yu2023:2211.16794} use Eulerian perturbation theory in combination with a halo model calibrated on N-body simulations to model the power spectrum multipoles from BOSS DR12 up to $k = 0.2 \hMpc$, finding $f\sigma_8(z = 0.61) = 0.455 \pm 0.026$, which is in excellent agreement with our constraint, albeit being at an effective redshift slightly higher than our sample.

Comparing with a full-shape analysis based on the Effective Field Theory of Large-Scale Structures \citep[EFTofLSS,][]{Carrasco2012:1206.2926}, we find an $f\sigma_8$ value than is $1.7\sigma$ higher than the one derived by \cite{d'Amico2020:1909.05271}, who fit the power spectrum multipoles of BOSS DR12 galaxies up to $k = 0.2 \hMpc$ and report $f\sigma_8(z=0.55) = 0.399 \pm 0.031$, which is closer to the \cite{Zhai2023:2203.08999} estimate. Although our $\Omegam$ constraints agree to within 0.1$\sigma$ with \cite{d'Amico2020:1909.05271}, their predicted amplitude of the primordial power spectrum, $A_s$, which is $2.3\sigma$ lower than PL18, drives their derived $f\sigma_8$ to lower mean values.

Finally, \cite{Semenaite2022:2111.0315} presented a clustering analysis of BOSS and eBOSS data, using information of the full shape of the BOSS clustering wedges presented by \cite{Sanchez2016}, and the multipoles from eBOSS quasars from \cite{Hou2021:2007.08998}. They derive $\sigma_8 = 0.815 \pm 0.044$ and $\Omegam = 0.290^{+0.012}_{-0.014}$, which agree with our constraints at the 0.4$\sigma$ and 1.2$\sigma$ level, respectively.

\subsection{Extended-$\Lambda$CDM constraints}

\begin{table}
    \renewcommand{\arraystretch}{1.4}
    \centering
    \caption{Constraints on single-parameter extensions to the base-$\Lambda$CDM model from the BOSS DR12 CMASS data, using the combination of density-split clustering and the galaxy 2PCF. }
    \rowcolors{2}{white}{gray!15}
    \begin{tabular} {|c|c|c|}
    \hline
     Parameter &  best-fit & mean $\pm \sigma$\\
    \hline
    $\Neff $ & $3.0759$ & $3.02^{+0.24}_{-0.27}$\\
    ${\rm d}n_s/{\rm d}\ln k$ & $-0.0022$ &$-0.005\pm 0.013$\\
    {$w_0$} & -0.9444 & $-0.956^{+0.046}_{-0.041}$\\
    \hline
    \end{tabular}
    \label{tab:extended_lcdm_constraints}
\end{table}

To explore potential deviations from $\Lambda$CDM, we have analyzed a grid of well-motivated extensions to the base model. 

Figure~\ref{fig:extended_lcdm} and Table~\ref{tab:extended_lcdm_constraints} summarize the results of single-parameter extensions to the base $\Lambda$CDM model. All these fits have been run using the baseline configuration of data vectors, scale cuts and model prescription, except for the addition of a single parameter to the cosmological model, which are incorporated one at a time. We do not find compelling proof supporting any of these extensions, as the marginalized posteriors for the additional parameters generally overlap with the base model within one standard deviation.

\subsubsection{Running of the spectral index}

\begin{figure*}
    \centering
    \begin{tabular}{c @{\hspace{1.0\tabcolsep}} c @{\hspace{0.5\tabcolsep}} c}
       \includegraphics[width=0.32\textwidth]{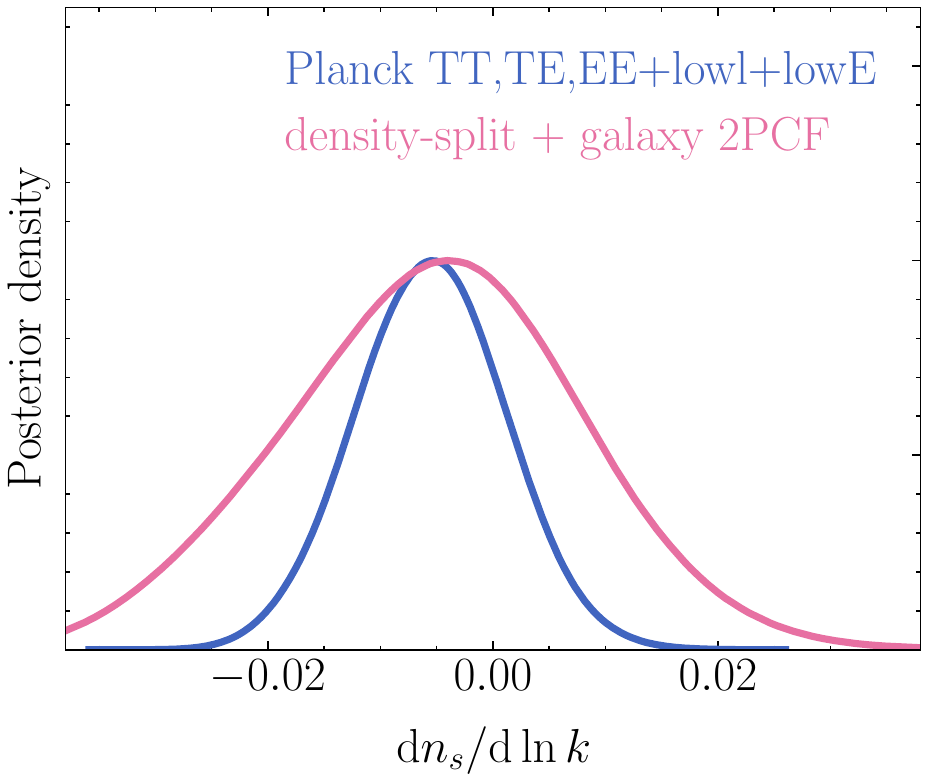}  & \includegraphics[width=0.30\textwidth]{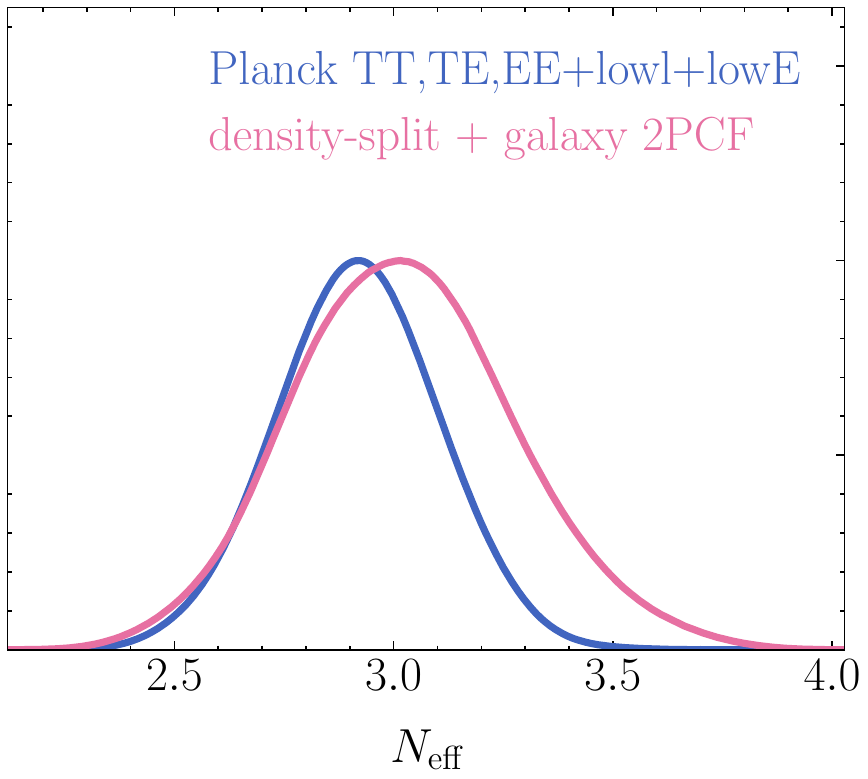} & \includegraphics[width=0.30\textwidth]{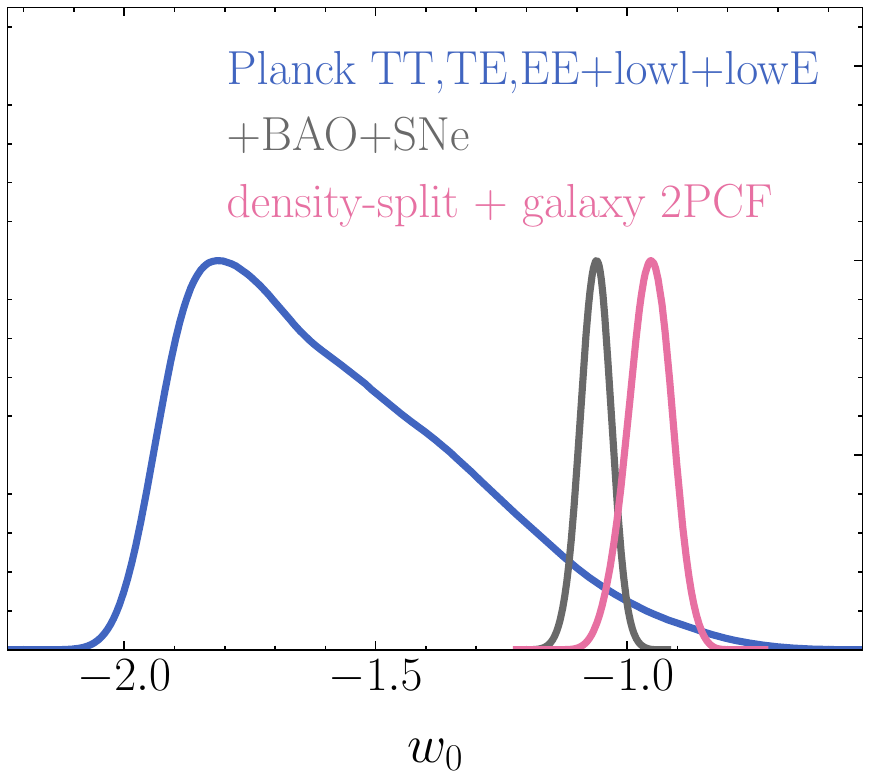}
    \end{tabular}
    \caption{Constraints on single-parameter extensions to the base-$\Lambda$CDM model from the galaxy 2PCF + density-split clustering fits on BOSS DR12 CMASS (pink), Planck TT,TE,EE+lowl+lowE (blue), and Planck TT,TE,EE+lowl+lowE+BAO+SNe (grey). Extensions include variations in the running of the spectral index of the primordial power spectrum, $\nrun$, the effective number of relativistic species, $\Neff$, and the dark energy equation of state parameter, $w_0$. \href{https://github.com/florpi/sunbird/blob/main/paper_figures/boss/cosmo_inference_cmass_nrun.py}{\faGithub}}
    \label{fig:extended_lcdm}
\end{figure*}

The first extension we look at is in regard to the scale dependence of the primordial density fluctuations. Here we characterize the primordial power spectrum as a power law with a normalization amplitude $A_s$, a spectral index $n_s$ and its first derivative with respect to $\ln k$ (also known as \textit{the running}):
\begin{align}
    P(k) &= A_s \left( \frac{k}{k_0} \right)^{n(k)}\\
    n(k) &= n_s - 1 + (1/2) (dn_s / d \ln k) \ln(k/k_0)\,,
\end{align}
where $k_0$ is a pivot wavenumber used to specify the point at which the power spectrum is normalized. Our base-$\Lambda$CDM constraints from Sect.~\ref{tab:base_lcdm_constraints} have assumed a zero running for the spectral index, finding $n_s = 0.970\pm 0.018$. CMB experiments also favour $n_s < 1$, which is predicted by common single-field slow-roll inflationary models \citep{Mukhanov2007}. Such models also predict a very small running, since it is only second-order in inflationary slow-roll parameters \citep{Kosowsky1995:astro-ph/9504071}, but it is possible to construct valid models that predict larger values. We find $\nrun = -0.005\pm 0.013$, which is consistent with zero running. The 1D marginalized posterior is shown in the left-hand side panel of Fig.~\ref{fig:extended_lcdm}, where we also overplot the constrain from the \textsc{Planck\_TTTEEE\_lowl\_lowE} likelihood from PL18, who find $-0.0055\pm 0.0067$. We warn the reader that the parameter range for the running shown in Fig.~\ref{fig:extended_lcdm} corresponds to our full prior range, beyond which we do not have simulations to sample the parameter space with our model. The posterior distribution quickly approaches zero near the prior walls, giving us confidence that our $1\sigma$ limits are not prior-dominated. However, we cannot confidently provide $3\sigma$ limits given this restriction.

\subsubsection{Effective number of relativistic species}

A relic neutrino background is a generic prediction of the standard hot Big Bang model \citep{Lesgourgues2014:1404.1740}. The constraints on the properties of relic neutrinos and other relativistic species beyond the Standard Model of particle physics is of special interest for large scale structure analyses. The combination of CMB experiments, galaxy and supernova surveys have put the tightest upper limits on the sum of neutrino masses \citep{Emas2019}, whereas Planck has constrained the density of light relics with sub-percent level precision \citep{Planck2020}.

In the instantaneous neutrino decoupling
limit, the density of radiation in the Universe (besides photons) can be written as \citep{Lesgourgues2014:1404.1740}:
\begin{equation}
    \frac{\rho_{\nu}}{\rho_{\gamma}} = \frac{7}{8}\Neff \left( \frac{4}{11}\right)^{4/3} \,,
\end{equation}
where $\Neff = 3$, usually called the effective number of relativistic species, is a convenient parametrization of the relativistic energy density of the Universe beyond just photons, in units of a single neutrino. Detailed calculations that go beyond the instantaneous neutrino decoupling limit, including neutrino oscillations, predict $\Neff \approx 3.046$ \citep{deSalas2016:1606.06986}.

For the base-$\Lambda$CDM constraints from Sect.~\ref{tab:base_lcdm_constraints}, we fixed $\Neff$ to the baseline value of 3.046. In this section, we vary this parameter during the inference analysis, finding $\Neff = 3.02^{+0.24}_{-0.27}$, which is a 8.4 per cent precision constraint, in excellent agreement with the PL18 measurement.

\subsubsection{Dark energy equation of state}

One of the main goals of modern observational cosmology is elucidating the nature of the accelerated expansion of the Universe. In the base-$\Lambda$CDM model, dark energy takes the form of a cosmological constant that has an equation of state $w_0 \equiv p / \rho = -1$, where $p$ and $\rho$ represent the pressure and density of the fluid. In this section, we explore models with a constant $w$ by letting it vary as a free parameter in the fit.

The right-hand side panel of Fig.~\ref{fig:extended_lcdm} shows the marginalized posterior of $w$ for the combination of the galaxy 2PCF and DSC. We find $ w_0 = -0.956^{+0.046}_{-0.041}$, which is consistent with a fiducial cosmological constant at the $1\sigma$ level. CMB data alone (blue contour) does not put a very tight constraint on $w_0$, as it is a $z \approx 1100$ measurement. The grey contours show the posterior resulting from the combination of the \textsc{Planck\_TTTEEE\_lowl\_lowE} PL18 likelihood with late-time probes of the expansion rate, including BAO measurements from BOSS DR12 \citep{Alam2017:1607.03155}, SDSS-MGS \citep{Ross2015:1409.3242}, and 6dFGS \citep{Beutler2011:1106.3366}, as well as type Ia supernovae distance measurements from \textsc{Pantheon} sample \citep{Scolnic2015:1508.05361}, and local estimates of the Hubble parameter from Milky Way Cepheid variables from \cite{Riess:2018:1801.01120}. This combination tightens the marginalized posterior to $w_0 = -1.041^{+0.060}_{-0.053}$, which agrees with our constraint at the $1.1\sigma$ level.

We emphasize that for our results from galaxy clustering, we have adopted the prior constraint for the acoustic scale $\theta_*$ from PL18, so our constraints on $w_0$ are not fully independent from the \textsc{Planck\_TTTEEE\_lowl\_lowE} likelihood.

\subsection{Test for systematics}
\label{subsec:systematics}

In this section, we carry out tests on mock galaxy catalalogues to look for systematics in the theorerical modelling, and we assess the robustness of our cosmological constraints to different choices in our inference pipeline.

\subsubsection{Recovery tests on the Nseries mocks}

We begin by testing our pipeline on the Nseries mock galaxy catalogues, which were calibrated onto the clustering of the CMASS NGC galaxy sample, matching its footprint and radial selection (Sect.~\ref{subsubsec:nseries}). We measure the galaxy 2PCF and density-split multipoles from each of the 84 mock realizations, and analyze each mock using the baseline configuration of our pipeline as we did in Sect.~\ref{tab:base_lcdm_constraints}. We estimate the covariance matrix of the data vectors from the MD-Patchy mocks, following the same procedure as described in Eq.~(\ref{eq:cov_data}).

The left panel of Fig.~\ref{fig:cosmo_inference_nseries} shows the distribution of recovered best-fit values from the 84 fits. The true cosmology of the simulations, which is shown by the vertical red lines, is well within the 68 per cent confidence region of the distribution, showing that our clustering pipeline is able to recover unbiased cosmological constraints even in the presence of complex survey masks, fiber collisions and non-uniform radial selection functions.

As a complementary test, we proceed to average the data vectors across the 84 mock realizations, and perform cosmological fits on the mean data vectors. The right panel of Fig.~\ref{fig:cosmo_inference_nseries} shows the recovered cosmological parameters in this setup, using two different covariance matrices. The grey contours show results with the usual covariance matched to the volume covered by the CMASS sample used throughout the paper. The black contours show results where the covariance is rescaled to match the comoving volume covered by 84 realizations\footnote{It should be noted however that the 84 realizations from Nseries are not fully independent from each other, since they were generated from only seven large cubic boxes that were rotated and trimmed in different ways to construct the cutsky mocks.} of the Nseries suite, which amounts to roughly $120\, (h^{-3}{\rm Gpc}^3)$, after applying the redshift cuts to match our data sample, $0.45 < z < 0.6$. Even using this large volume, the true cosmology of the simulations falls within one standard deviation of the marginalized parameter posterior distributions, highlighting the robustness of our pipeline even for datasets far larger than the one analysed in this paper.

\begin{figure}
    \centering
    \includegraphics[width=0.9\columnwidth]{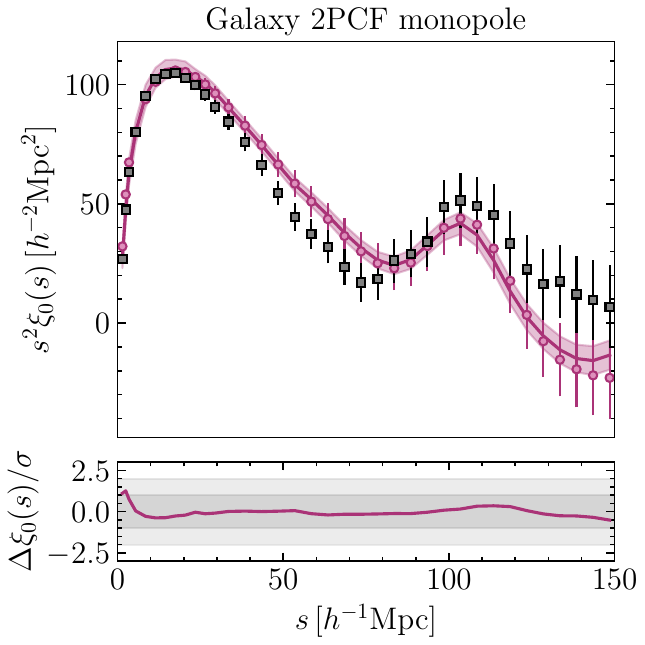}
    \caption{Monopole of the galaxy two-point correlation function, averaged over 84 realizations of the Nseries mocks (violet circles with error bars), along with the best-fit model and its associated uncertainty (violet solid line and bands). Also shown is the  monopole from the BOSS CMASS galaxy sample (grey squares with error bars). The lower sub-panel shows the difference between the Nseries mocks and the best-fit model, in units of the error bars. The dark-grey and grey shaded regions demarcate $1\sigma$ and $2\sigma$ offsets, respectively. \href{https://github.com/florpi/sunbird/blob/main/paper_figures/boss/multipoles_nseries.py}{\faGithub}}
    \label{fig:nseries_tpcf}
\end{figure}

Figure~\ref{fig:nseries_tpcf} shows the best fit to the galaxy 2PCF measured from the mean of the Nseries samples (using the covariance associated to a single CMASS volume), where we also overplot the measurement from the CMASS sample for comparison. Overall, the fit to the Nseries mocks is accurate at all scales, with the best-fit model always falling within one standard deviation of the error bars. This can be contrasted with the fit to the CMASS data seen earlier in Sect.~\ref{subsec:model_fits}, where the data measurement shows a lower and higher clustering amplitude at intermediate and large scales compared to the model, respectively. CMASS also shows a similar difference in clustering with respect to the mean of the Nseries mocks. The $\chi^2$ between the mocks and data monopole is 30 for 23 degrees of freedom, which corresponds roughly to a $2\sigma$ shift. Although we do not show the other statistics for brevity, we have observed the same trends for DSC. However, it is worth keeping in mind that the cosmology of the Nseries simulations differs substantially from the fiducial cosmology we adopted to convert redshifts to distances, so the Nseries multipoles could be affected by more severe AP distortions than the CMASS data, if the true cosmology of the Universe is closer to our fiducial cosmology compared to the mocks.

\begin{figure}
    \centering
    \includegraphics[width=0.9\columnwidth]{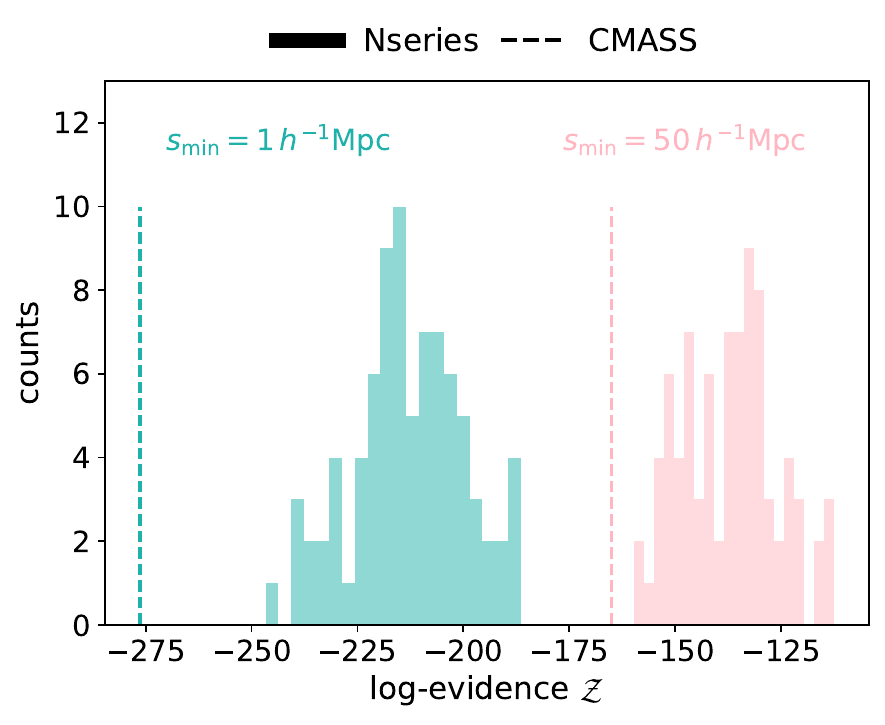}
    \caption{Logarithm of the Bayesian evidence obtained from individual fits to the Nseries mocks (histograms) and to the BOSS DR12 CMASS data (dashed lines). We show results for two different minimum scale cuts: $s_{\rm min} = 1 \Mpch$ (green) and $s_{\rm min} = 50 \Mpch$ (pink). The evidence of the CMASS fit is significantly lower than any of the Nseries mocks in both cases, but this discrepancy is reduced when excluding small scales. \href{https://github.com/florpi/sunbird/blob/main/paper_figures/boss/evidence.py}{\faGithub}}
    \label{fig:evidence}
\end{figure}

Figure~\ref{fig:evidence} shows the distribution of log-evidence from the fits to the 84 Nseries mocks and CMASS. We include the results obtained using two different scale cuts: $s_{\rm min} = 1 \Mpch$ and $s_{\rm min} = 50 \Mpch$. We see that regardless of the scale cut that is used, the evidence of the CMASS fit is significantly lower than for any of the Nseries mocks. This means that, overall, the model is a much better description of the mocks than it is of the data, even when marginalizing over all the parameter space of the model. This supports the hypothesis that there might be residual systematic effects in the CMASS clustering catalogues that are currently unaccounted for by the weighting procedure  \citep{Lavaux2019:1909.06396}. While removing all the information below $s = 50 \Mpch$ reduces the gap between Nseries and CMASS, it is still statistically unlikely that the large offsets between the baseline model and the data can be fully explained by a random fluctuation due to sample variance from the observations, if we assume that the mocks are a faithful reproduction of the CMASS data. This could be a combination of the above-mentioned systematic effects with insufficient models of the halo-galaxy connection on small scales. Given the small number of Nseries mocks, we cannot translate these findings into more quantitative statements.

\subsubsection{Scale cuts}

\begin{figure}
    \centering \includegraphics[width=0.9\columnwidth]{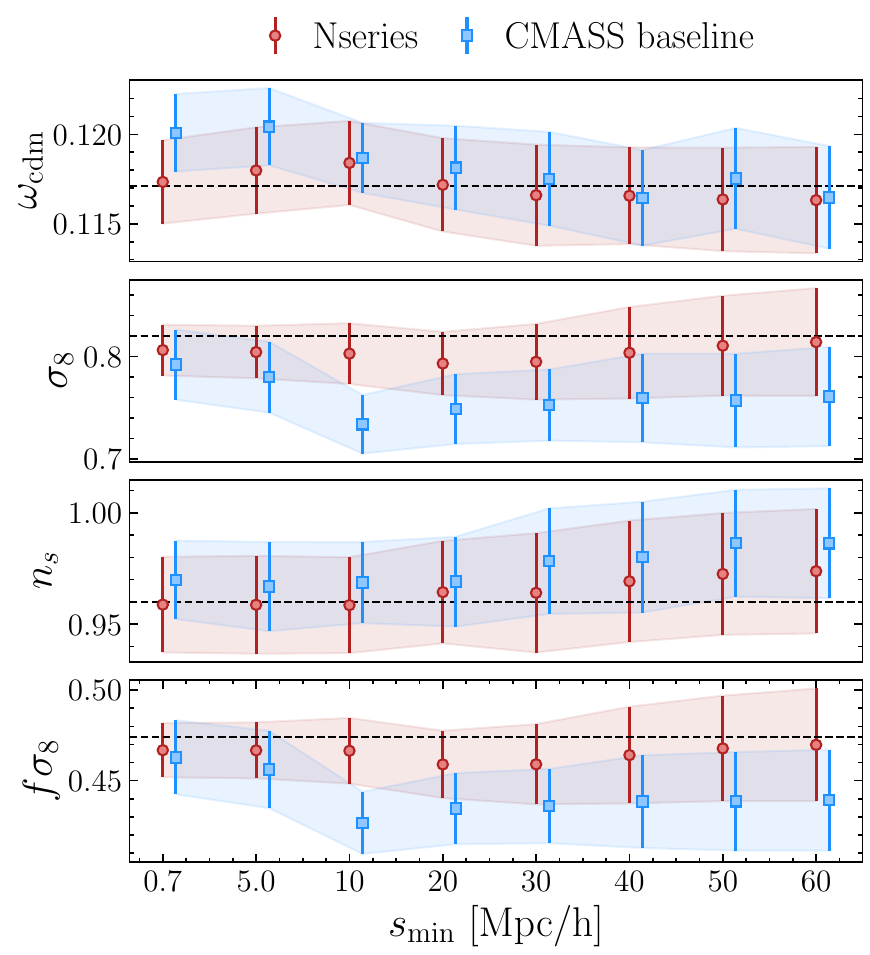}
    \caption{Impact of the minimum scale cut on the constraints on base-$\Lambda$CDM parameters and $f\sigma_8$ when fitting the mean of the Nseries mocks (red circles), or the BOSS DR12 CMASS data blue squares). The horizontal dashed line shows the true cosmology of the Nseries simulations. Note that the horizontal axes do not use a linear scale. \href{https://github.com/florpi/sunbird/blob/main/paper_figures/boss/scalecuts.py}{\faGithub}}
    \label{fig:scalecuts_nseries}
\end{figure}

Another important aspect we study with the Nseries mocks is the choice of scales that are used in the main data analysis. Figure~\ref{fig:scalecuts_nseries} shows the marginalized constraints on $\omegac$, $\sigma_8$, $n_s$, and $f\sigma_8$, adopting eight different minimum scale cuts, ranging from $1 \Mpch$ to $60 \Mpch$. For these measurements, we use the mean of the 84 Nseries mocks as the data vector, and a covariance associated to a single CMASS volume. We observe that for Nseries, the mean values of the marginalized posteriors are very stable when changing the scale cuts for all parameters that are considered. It is interesting to note that the size of the error bars does not shrink beyond $\approx 20 \Mpch$. There are two main reasons for this. The first one is the inclusion of the model uncertainty in the covariance matrix that is used for the calculation of the likelihood [Eq.~(\ref{eq:total_covariance})]. At these scales, although the emulator error has a percent-level accuracy, it starts to dominate the total error budget for the monopole of the 2PCF and the density-split CCF. This ensures that the cosmological inference is always robust, even when including scales where the emulator is less accurate than the precision of the data, which comes at the expense of not being able to extract all the information there is available. The second reason is that even though we are imposing a minimum scale cut in the multipoles, the DSC quintiles are defined using small-scale information from the density field, which also propagates into the multipoles at larger scales, as shown in \cite{Paillas2022:2209.04310}.

Figure~\ref{fig:scalecuts_nseries} also shows how the constraints from CMASS data change depending on the minimum scale cut. Although the results we get using $s_{\rm min} = 1 \Mpch$ are consistent with those obtained with more conservative cuts to within 1$\sigma$, we observe some interesting trends with scale. $\omegac$ shows some tendency towards larger mean values the more small-scale information we include. Furthermore, $\sigma_8$ transitions to larger mean values going from $s_{\rm min} = 20 \Mpch$ to $s_{\rm min} = 5 \Mpch$. The fact that these two effects are not observed in the Nseries mocks could be partially attributed to noise in the CMASS data vector, residual observational systematic effects that the mocks do not capture, or misunderstandings in the galaxy-halo connection modelling, which is mostly constrained by small scales. One of the most common systematics that can affect small-scale clustering is fiber assignment, which artificially lowers the clustering amplitude on scales smaller than the fiber collision angular scale. The Nseries mocks are already infused with fiber collisions that should closely match the BOSS data, so any important effect coming from this should also be imprinted in the multipoles measured from the mocks.

Based on the fact that the constraints from Nseries are robust against variations in $s_{\rm min}$, and that the CMASS constraints with different scale cuts are consistent to within 1$\sigma$, we adopt $s_{\rm min} = 1 \Mpch$ as the baseline for our analysis.

\subsubsection{Systematic error budget}

\begin{figure*}
    \centering
    \begin{tabular}{cc}
      \includegraphics[width=0.45\textwidth]{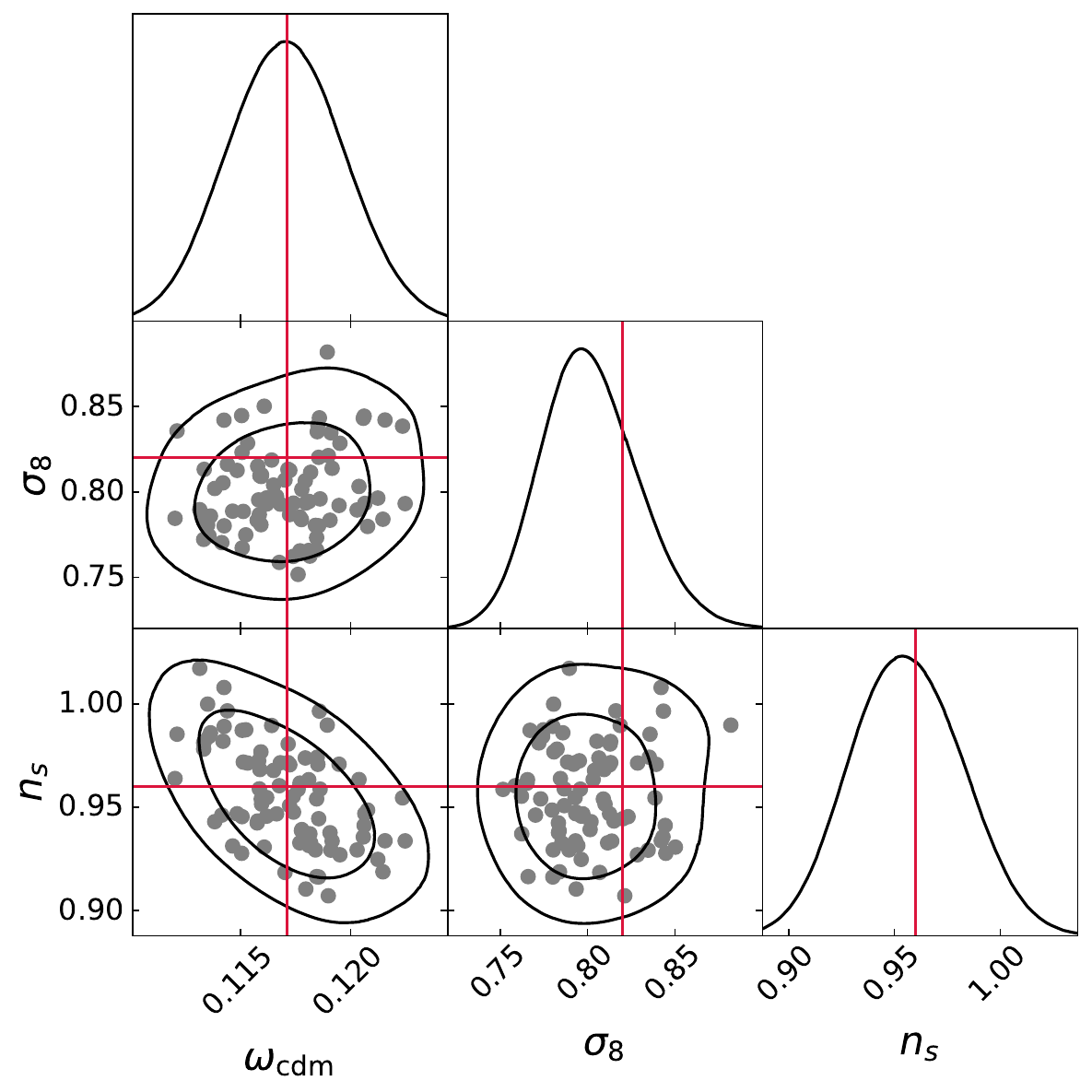}  & \includegraphics[width=0.45\textwidth]{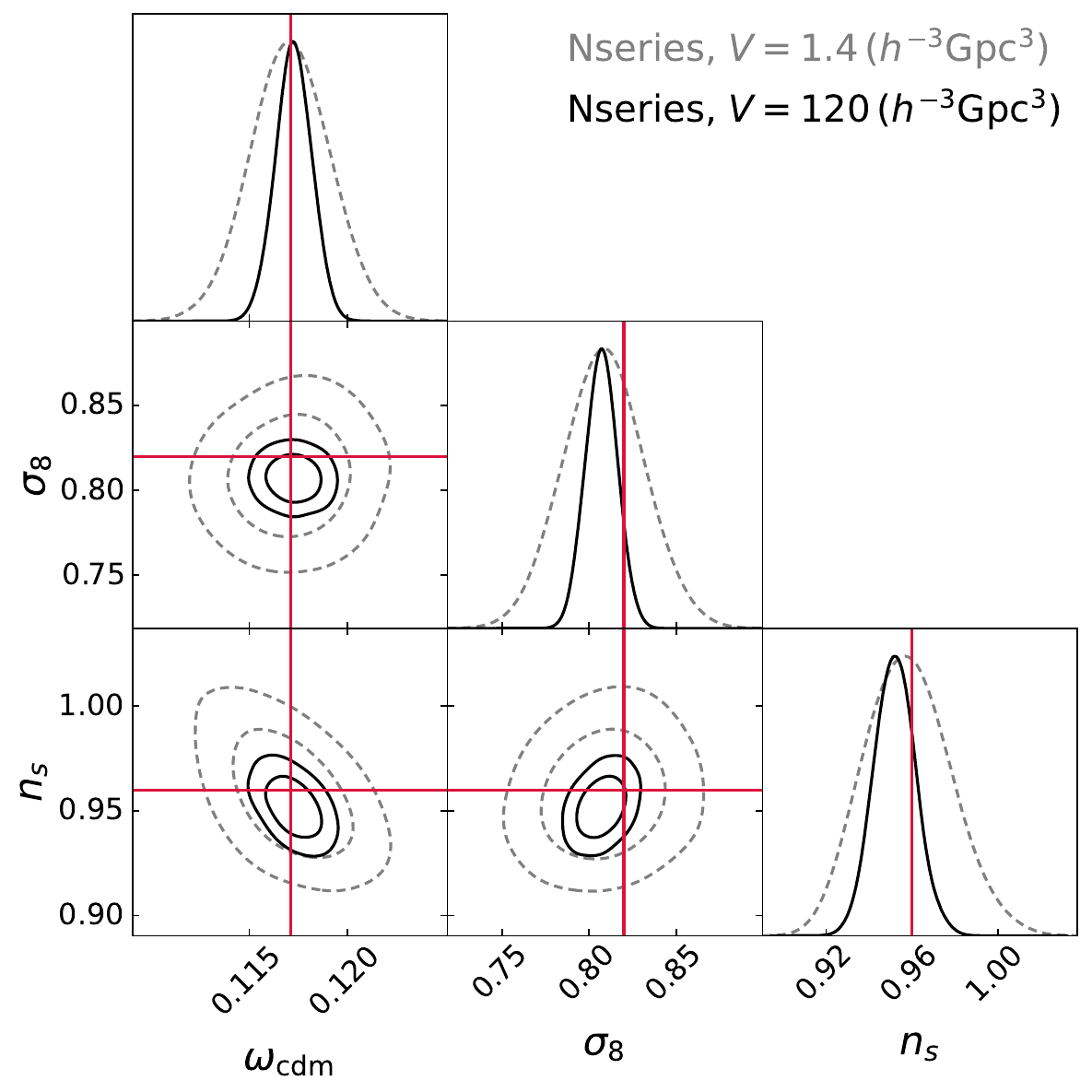}
    \end{tabular}
    \caption{Marginalized constraints on $\omega_{\rm cdm}$, $\sigma_8$, and $n_s$ from fits to the Nseries mock catalogues, which were calibrated onto the clustering of the CMASS NGC sample, matching its number density and geometry. The fits were performed using the baseline configuration of our analysis, consisting on the combination of the density-split and the galaxy correlation functions at scales $1.0\, \Mpch < s < 150\, \Mpch$.\\
    Left: the dots represent the best-fit values of individual fits to 84 realizations of the Nseries mocks. The contours show the 68 and 85 per cent confidence regions of the distribution of individual fits. Right: fits to data vectors that are averaged over the 84 Nseries mocks, using a covariance rescaled to match the volume of the CMASS sample used in our main analysis (dashed-grey) or the total volume of the Nseries suite (solid-black). \href{https://github.com/florpi/sunbird/blob/main/paper_figures/boss/cosmo_inference_nseries.py}{\faGithub}}
    \label{fig:cosmo_inference_nseries}
\end{figure*}

Based on the tests described in the previous section, we use the Nseries mocks to determine the contribution to the systematic error budget coming from our emulator. This was trained on periodic boxes at a fixed redshift, but we are applying it to fit a survey that has a non-uniform footprint and radial selection, along with fiber collision effects that artificially decrease the clustering on small scales. Thus, there is the possibility of errors in addition to the emulator error term determined from the test sample when constructing the emulator.

To look for such error terms, we calculate the offset between the expected value of each cosmological parameter and the mean of the marginalized posteriors from the fit to the mean of 84 mocks, using the covariance of a $120 (h^{-3}{\rm Gpc}^3)$ volume (solid contours from the right-hand side panel of Fig.~\ref{fig:cosmo_inference_nseries}). For the combination of DSC and the galaxy 2PCF, the offsets and their associated $2\sigma$ uncertainties are:
\begin{alignat*}{2}
    &\Delta \omegac &&= 0.00015^{+0.0018}_{-0.0018}\\
    &\Delta \sigma_8 &&= 0.01288^{+0.018}_{-0.018}\\
    &\Delta n_s &&= 0.00801^{+0.020}_{-0.019}\\
    &\Delta f\sigma_8 &&= 0.00657^{+0.011}_{-0.011} \quad \text{(DSC + galaxy 2PCF)}\,, 
\end{alignat*}
while the fits that only include the galaxy 2PCF give
\begin{alignat*}{2}
    &\Delta \omegac &&= 0.00002^{+0.0040}_{-0.0039}\\
    &\Delta \sigma_8 &&= 0.01151^{+0.046}_{-0.045}\\
    &\Delta n_s &&= 0.00212^{+0.039}_{-0.038}\\
    &\Delta f\sigma_8 &&= 0.00607^{+0.028}_{-0.027} \quad \text{(galaxy 2PCF)}\,. 
\end{alignat*}
We do not find statistically significant offsets from the expected values, with the shifts for all parameters appearing well within their $2\sigma$ limits. However, the level of precision to which we can carry out this test is set by the number of mocks that are available, which limits the total effective volume used in the test. We take the $2\sigma$ limit of the marginalized distributions of best-fit values (left-hand side of Fig.~\ref{fig:cosmo_inference_nseries}), divided by $\sqrt{84}$, as a conservative (maximum) estimate of the systematic error for each parameter, which are then added in quadrature to the associated statistical errors. We estimate this from the distribution of individual fits divided by $\sqrt{84}$ rather than from the fit to the mean, because the $2\sigma$ limits of the latter are dominated by the emulator error on the multipoles, which is already included in the covariance when we calculate the likelihood [Eq.~(\ref{eq:total_covariance})]. In this way, we avoid double counting the emulator uncertainty, only including the limiting precision that comes from the finite number of mocks used for the test. With this setup, the systematic error added to each parameter for the combination of DSC + galaxy 2PCF combination is:
\begin{alignat*}{2}
    &\sigma^{\rm sys}_{\omegab}  &&= 0.0000156\\
    &\sigma^{\rm sys}_{\omegac}  &&=  0.00037\\
    &\sigma^{\rm sys}_{\sigma_8} &&= 0.00370 \\
    &\sigma^{\rm sys}_{n_s}      &&= 0.00354  \quad \text{(DSC + galaxy 2PCF)}\,,
\end{alignat*}
while for the 2PCF-only fits we get
\begin{alignat*}{2}
    &\sigma^{\rm sys}_{\omegab}  &&= 0.0000074\\
    &\sigma^{\rm sys}_{\omegac}  &&= 0.00046\\
    &\sigma^{\rm sys}_{\sigma_8} &&= 0.00519 \\
    &\sigma^{\rm sys}_{n_s}      &&= 0.00348  \quad \text{(galaxy 2PCF)}\,.
\end{alignat*}
These systematic errors are added in quadrature to the statistical error budget. The systematic errors are negligible compared to the statistical errors on the parameter constraints from CMASS, so the values reported in Table~\ref{tab:base_lcdm_constraints} are unaffected, up to the significant figures that are shown.

\subsubsection{Robustness against pipeline settings}

Here we explore the robustness of our clustering analysis against different choices of settings in the inference pipeline. The tests on this section are performed on the real data, and no longer using the Nseries mocks. As a reminder, our baseline configuration consists of:
\begin{itemize}
    \item[-] Data vector: Concatenation of the monopole and quadrupole moments of the DSC auto-correlation and cross-correlation functions, using four quintiles ($\Q_0$, $\Q_1$, $\Q_3$, and $\Q_4$) and the galaxy 2PCF.
    \item[-] Parameter space: Base-$\Lambda$CDM model with an extended HOD framework, including velocity bias and environment-based assembly bias:
    \begin{alignat*}{2}
        \qquad &\bm{\theta}_{\rm cosmo} &&= \{ \omegac, \omegab, \sigma_8, n_s \}\\
        \qquad &\bm{\theta}_{\rm HOD} &&= \{ M_{\rm cut}, M_1, \sigma, \alpha, \kappa, \alpha_{\rm vel, c}, \alpha_{\rm vel, s}, B_{\rm cen}, B_{\rm sat} \}
    \end{alignat*}
    \item[-] Priors: Uniform priors for all parameters except the baryon density, for which we adopt a BBN-like Gaussian prior, as detailed in Table~\ref{tab:priors}.
    \item[-] Error budget: The covariance used in the likelihood calculation includes contributions from sample variance of the data vector, and model uncertainty associated with the emulator training.
\end{itemize}

\begin{figure*}
    \centering
    \includegraphics[width=0.7\textwidth]{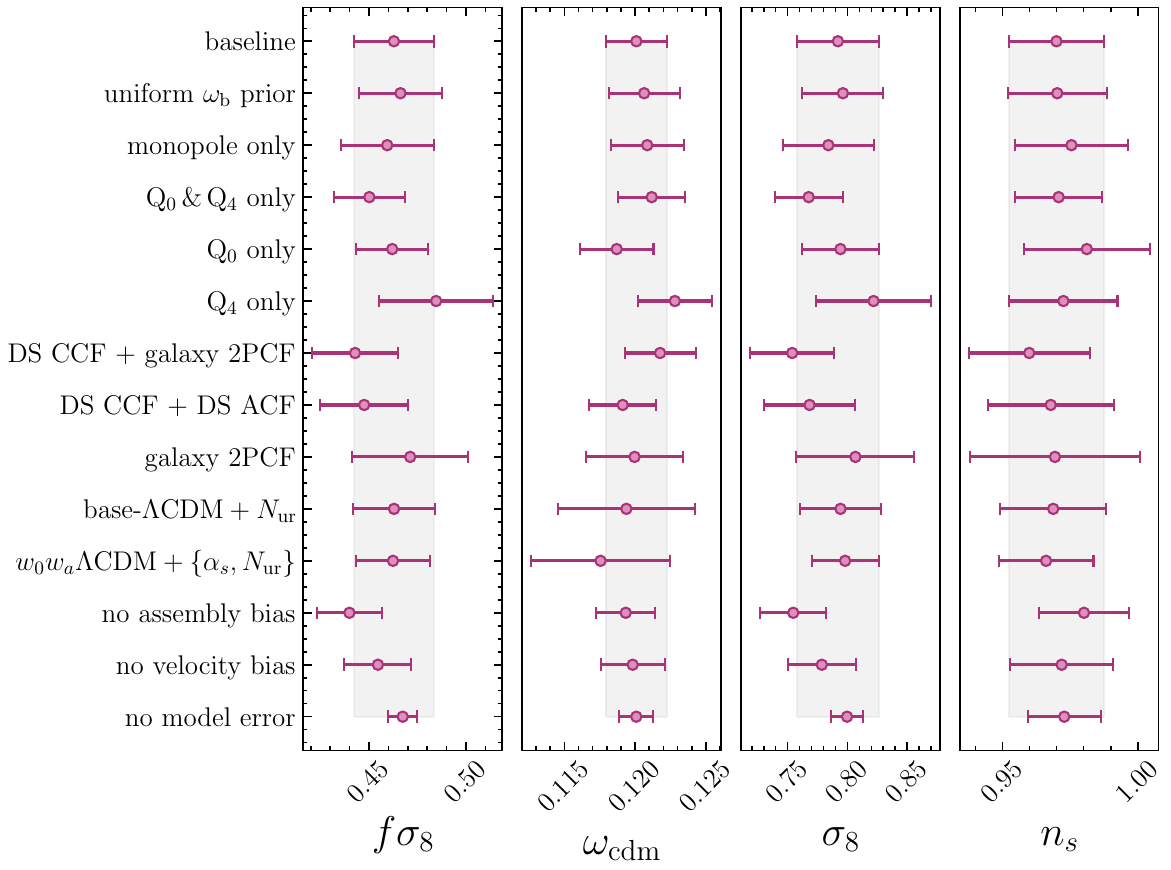}
    \caption{A comparison of the constraints on the base-$\Lambda$CDM parameters and the derived $f\sigma_8$ value from fits run with different analysis settings. Our baseline analysis, shown at the top, was run by simultaneously fitting the galaxy 2PCF and density-split multipoles with a base-$\Lambda$CDM + extended-HOD model, using scales $1 \Mpch < s < 150 \Mpch$. The other points represent variations to that baseline configuration, either by changing the data vector or the model prescription. \href{https://github.com/florpi/sunbird/blob/main/paper_figures/boss/whisker_cmass.py}{\faGithub}}
    \label{fig:whisker_cmass}
\end{figure*}

Figure~\ref{fig:whisker_cmass} shows the constraints that result from varying various aspects of these settings. Starting from top to bottom, we try a uniform prior for $\omega_b$, finding that it shifts the means of the marginalized posteriors by less than $0.2\sigma$, but degrading the constrainig power on $\omegac$ by 15 per cent. $\omega_b$ itself is basically unconstrained with our analysis under a uniform prior, which is the main motivation for adopting a BBN prior, following other clustering studies in the literature \citep[e.g., ][]{Alam2017:1607.03155, Ivanov2020:1909.05277, Philcox2021a:2112.04515}.

Using only the monopole of the correlation functions weakens the precision of the constraints on $\omegac$ and $\sigma_8$ by 18 and 13 per cent, respectively, resulting in a 18 per cent degradation of the precision on $f\sigma_8$.

Using only the most underdense and overdense quintiles ($\Q_0$ and $\Q_4$, respectively) degrades the precision on $\omegac$ by 9 per cent. The precision on $\sigma_8$ increases by 15 per cent and its mean is shifted to slightly lower values. The increase in precision for $\sigma_8$ might seem counter intuitive, but can be explained by the fact that individual quintiles can sometimes lead to marginalized posteriors with slightly different mean values, which is exemplified by the rows where we fit $\Q_0$ and $\Q_4$ separately. This can produce wider contours when all quintiles are fitted simultaneously. Overall, the fact that the constraining power on most parameters is not severely degraded when we fit only the extreme quintiles (which effectively discards half of the DSC data set) agrees with the picture that most of the information from DSC comes from the very under-dense and over-dense regions, as suggested by \cite{Paillas2021}.

DSC by itself predicts $\sigma_8 = 0.768^{+0.037}_{-0.043}$, which is slightly lower than the predicted value from the 2PCF-only fit. However, both the DSC-only and 2PCF-only fits are consistent within one standard deviation with the baseline fit.

We observe that the precision on $\omegac$ degrades by a factor of 2.4 when we allow $\Neff$ to vary, which is due to the strong correlation between these parameters. However, the precision for the other parameters remains relatively stable. If we also allow $\nrun$ and $w_0$ to vary, it does not significantly affect the constraining power on the base $\Lambda$CDM parameters.

We find an interesting shift of the estimated $\sigma_8$ to lower values when we fix the assembly bias or velocity bias parameters during the fit. As shown in Appendix~\ref{ap:hod_constraints}, our best-fit model constrains the environment-based assembly bias parameters to be negative, meaning that galaxies preferentially form in haloes in denser environments. We observe a mild correlation between $\sigma_8$ and $\alpha$, $\alpha_{{\rm vel},s}$ and $B_{\rm sat}$ that is likely to explain the shifts in $\sigma_8$ towards lower values when removing assembly or velocity bias parameters.

Finally, we run a fit neglecting the contribution of the model uncertainty to the error budget, which can be achieved by removing ${\bf C}^{\rm emu}$ and ${\bf C}^{\rm sim}$ from the covariance used to calculate the likelihood. This has a drastic impact on the derived constraints, resulting in a precision that can be more than two times better than the baseline analysis. Although for this particular test we observe that the mean values of the recovered parameters do not shift significantly with respect to the baseline fit, we have explicitly verified that removing the emulator error results in cosmological constraints that can be biased at a more than 3$\sigma$ level when fitting the mean of the Nseries mocks down to $1 \Mpch$.

\section{Discussion and conclusions}
\label{sec:conclusions}

We have presented a clustering analysis of the DR12 BOSS CMASS galaxy sample at $0.45 < z < 0.6$, using simulation-based models of the galaxy two-point correlation function (2PCF) and density-split clustering (DSC). Our theory framework, which is presented in detail in our companion paper \citep{Cuesta-Lazaro2023:2309.16539}, is based on emulators trained on high-fidelity mock galaxy catalogues, which forward model the cosmological dependence of the full shape of the galaxy 2PCF and DSC multipoles, including redshift-space and Alcock-Paczynski distortions. 

It should be noted that due to the limitations of the simulation data available for training the emulator, our model fits impose a fixed prior on the acoustic scale $\theta_*$ measured from the CMB \citep{Planck2020}. This is, however, a very precise and model-independent measurement, so this prior does not significantly restrict our conclusions about the models analysed.

We have validated our theory model against the Nseries mock galaxy catalogues, which were calibrated onto the clustering and selection properties of the CMASS galaxy sample, finding that we can recover unbiased cosmological constraints even using a volume of $120 (h^{-3}{\rm Gpc}^3)$, which is 84 times larger than the volume examined in our data analysis.

For our base $\Lambda$CDM analysis, we find that the galaxy 2PCF constrains $\omegac$, $\sigma_8$, and $n_s$ with a precision of 2.8, 6.1, and 3.2 per cent, respectively, using a scale range $1 \Mpch < s < 150 \Mpch$. Adding the DSC multipoles using the same scale range tightens the constraining power to a precision of 1.8, 4.3 and 1.8 per cent, respectively, obtaining $\omegac = 0.1201\pm 0.0022$, $\sigma_8 = 0.792\pm 0.034$, and $n_s = 0.970\pm 0.018$. This is a factor of 1.6, 1.4, and 1.8 of improvement in precision with respect to the 2PCF-only constraints, respectively.

Combining the galaxy 2PCF and DSC multipoles, we derive $f \sigma_8 = 0.462\pm 0.020$ at $z \approx 0.525$, which is a 4.3 per cent constraint. In comparison, the main BOSS clustering analysis presented in \cite{Alam2017:1607.03155} derived a 8.3 per cent constraint on $f\sigma_8$ using both Galatic caps for their $0.4 < z < 0.6$ redshift bin. We obtain 1.9 times better precision using only the Northern Galactic cap and a narrower redshift bin. This improvement mainly comes from the inclusion of higher-order clustering information that is captured by DSC, and the addition of non-linear scales in the fitting. Our $f \sigma_8$ constraint is largely consistent with Planck 2018 base-$\Lambda$CDM predictions \citep{Planck2020}, and also agrees well with other clustering studies in the literature that use the same galaxy sample \citep{Yuan2022b:2203.11963, Yu2023:2211.16794}. Our base-$\Lambda$CDM cosmological constraints are summarized in Table~\ref{tab:base_lcdm_constraints}.

We have also performed fits with single-parameter extensions to base-$\Lambda$CDM, where we vary the running of the spectral index of the primordial power spectrum ($\nrun$), the density of massless relic neutrinos ($\Neff$), and the dark energy equation of state parameter $w_0$. Overall, we do not find compelling evidence for deviations from the base $\Lambda$CDM model, obtaining $\Neff = 3.02_{-0.27}^{+0.24}$, $\nrun = -0.005 \pm 0.013$, and $w_0 = -0.956_{-0.041}^{+0.046}$.

We have used an extended halo occupation distribution (HOD) framework to model the LRG galaxy-halo connection, using halo catalogues from the \textsc{AbacusSummit} suite of simulations. Our HOD constraints are largely consistent with the findings of \cite{Yuan2022:2110.11412} and \cite{Yuan2022b:2203.11963}. We constrain the minimum halo mass for hosting centrals to be $\log M_{\rm cut} = 12.65^{+0.08}_{-0.11}$, and the typical halo mass for hosting one satellite $\log M_1 = 13.69^{+0.10}_{-0.15}$. We find signs of environment-based assembly bias, suggesting a preference for galaxies to form in haloes around denser environments.

Overall, we find that our model is able to fit the Nseries mock galaxy catalogues much more accurately than the CMASS data itself. We deem that this is not a problem specific to our emulation framework, since the mean clustering signal of the Nseries mocks, which agrees with the clustering of the AbacusSummit simulations for similar cosmologies, shows a similar offset with respect to the CMASS data. This suggests the possibility that there might be residual systematic effects in the data that are currently not captured by the mocks and are not taken into account by the weighting procedure that we adopt when calculating the clustering.

It is also worth noting that our emulator has been trained on HOD catalogues at a fixed redshift, $z = 0.5$, and we do not include any possible effects of redshift dependence on the modelling. Furthermore, the effective redshift of our CMASS sample, $z_{\rm eff} = 0.525$, differs slightly from the redshift of the HOD catalogues. Although the relatively narrow redshift range we are imposing in the CMASS catalogue ($0.45 < z < 0.6$) could alleviate some concerns regarding the redshift dependence of the theoretical model, a careful study of the impact of this simplification on the parameter constraints is going to become even more relevant when extending our framework to other datasets such as eBOSS  \citep{Dawson2016}. We plan to study this in future work using the \textsc{AbacusSummit} lightcone simulations that have recently become available \citep{Hadzhiyska2022:2110.11413}.

On-going and upcoming galaxy redshift surveys, such as DESI \citep{desi}, Euclid \citep{euclid} and the Nancy Grace Roman Space Telescope \citep{roman} will open exciting avenues to use beyond-two-point statistics for cosmology, not only in terms of improved constraints on cosmological parameters, but also in regards to our understanding of the galaxy-halo connection, the treatment of observational systematics, and the possibility of finding surprises in the data that can challenge our preconceptions about the Cosmos.

\section*{Acknowledgements}

The authors thank Etienne Burtin for helpful discussions throughout the development of this project. This research was enabled in part by the support provided by Compute Ontario (computeontario.ca) and the Digital Research Alliance of Canada (alliancecan.ca). WP acknowledges the support of the Natural Sciences and Engineering Research Council of Canada (NSERC), [funding reference number RGPIN-2019-03908] and from the Canadian Space Agency. SN acknowledges support from an STFC Ernest Rutherford Fellowship, grant reference ST/T005009/2. FB is a University Research Fellow and has received funding from the European Research Council (ERC) under the European Union’s Horizon 2020 research and innovation program (grantagreement853291). YC acknowledges the support of the Royal Society through a University Research Fellowship.

Research at Perimeter Institute is supported in part by the Government of Canada through the Department of Innovation, Science and Economic Development Canada and by the Province of Ontario through the Ministry of Colleges and Universities. This work was supported by the U.S. Department of Energy through grant DE-SC0013718 and under DE-AC02-76SF00515 to SLAC National Accelerator Laboratory, and by the Kavli Institute for Particle Astrophysics and Cosmology. This project used resources of the National Energy Research Scientific Computing Center (NERSC), a U.S. Department of Energy Office of Science User Facility located at Lawrence Berkeley National Laboratory, operated under Contract No. DE-AC02-05CH11231. The \textsc{AbacusSummit} simulations were run at the Oak Ridge Leadership Computing Facility, which is a DOE Office of Science User Facility supported under Contract DE-AC05-00OR22725.

For the purpose of open access, the authors have applied a CC BY public copyright licence to any Author Accepted Manuscript version arising.

\section*{Data Availability Statement}

The source code to generate the figures in this manuscript is available at \url{https://github.com/florpi/sunbird}. The data from the \textsc{AbacusSummit} suite of simulations can be found at \url{https://abacusnbody.org}. Any other data necessary to reproduce the results of this work will be shared upon reasonable request to the corresponding author.


\bibliographystyle{mnras}
\bibliography{references} 

\begin{thebibliography}{}
\makeatletter
\relax
\def\mn@urlcharsother{\let\do\@makeother \do\$\do\&\do\#\do\^\do\_\do\%\do\~}
\def\mn@doi{\begingroup\mn@urlcharsother \@ifnextchar [ {\mn@doi@}
  {\mn@doi@[]}}
\def\mn@doi@[#1]#2{\def\@tempa{#1}\ifx\@tempa\@empty \href
  {http://dx.doi.org/#2} {doi:#2}\else \href {http://dx.doi.org/#2} {#1}\fi
  \endgroup}
\def\mn@eprint#1#2{\mn@eprint@#1:#2::\@nil}
\def\mn@eprint@arXiv#1{\href {http://arxiv.org/abs/#1} {{\tt arXiv:#1}}}
\def\mn@eprint@dblp#1{\href {http://dblp.uni-trier.de/rec/bibtex/#1.xml}
  {dblp:#1}}
\def\mn@eprint@#1:#2:#3:#4\@nil{\def\@tempa {#1}\def\@tempb {#2}\def\@tempc
  {#3}\ifx \@tempc \@empty \let \@tempc \@tempb \let \@tempb \@tempa \fi \ifx
  \@tempb \@empty \def\@tempb {arXiv}\fi \@ifundefined
  {mn@eprint@\@tempb}{\@tempb:\@tempc}{\expandafter \expandafter \csname
  mn@eprint@\@tempb\endcsname \expandafter{\@tempc}}}

\bibitem[\protect\citeauthoryear{{Alam} et~al.,}{{Alam}
  et~al.}{2017}]{Alam2017:1607.03155}
{Alam} S.,  et~al., 2017, \mn@doi [\mnras] {10.1093/mnras/stx721}, \href
  {https://ui.adsabs.harvard.edu/abs/2017MNRAS.470.2617A} {470, 2617}

\bibitem[\protect\citeauthoryear{{Alcock} \& {Paczynski}}{{Alcock} \&
  {Paczynski}}{1979}]{Alcock1979}
{Alcock} C.,  {Paczynski} B.,  1979, \mn@doi [\nat] {10.1038/281358a0}, \href
  {https://ui.adsabs.harvard.edu/abs/1979Natur.281..358A} {281, 358}

\bibitem[\protect\citeauthoryear{{Aung} et~al.,}{{Aung}
  et~al.}{2023}]{Aung2023:2209.12918}
{Aung} H.,  et~al., 2023, \mn@doi [\mnras] {10.1093/mnras/stac3514}, \href
  {https://ui.adsabs.harvard.edu/abs/2023MNRAS.519.1648A} {519, 1648}

\bibitem[\protect\citeauthoryear{Aver, Olive  \& Skillman}{Aver
  et~al.}{2015}]{Aver2015:1503.08146}
Aver E.,  Olive K.~A.,   Skillman E.~D.,  2015, \mn@doi [Journal of Cosmology
  and Astroparticle Physics] {10.1088/1475-7516/2015/07/011}, 2015, 011

\bibitem[\protect\citeauthoryear{{Banerjee} \& {Abel}}{{Banerjee} \&
  {Abel}}{2021}]{Banerjee2021}
{Banerjee} A.,  {Abel} T.,  2021, \mn@doi [\mnras] {10.1093/mnras/staa3604},
  \href {https://ui.adsabs.harvard.edu/abs/2021MNRAS.500.5479B} {500, 5479}

\bibitem[\protect\citeauthoryear{{Bautista} et~al.,}{{Bautista}
  et~al.}{2021}]{Bautista2021}
{Bautista} J.~E.,  et~al., 2021, \mn@doi [\mnras] {10.1093/mnras/staa2800},
  \href {https://ui.adsabs.harvard.edu/abs/2021MNRAS.500..736B} {500, 736}

\bibitem[\protect\citeauthoryear{{Beutler} et~al.,}{{Beutler}
  et~al.}{2011}]{Beutler2011:1106.3366}
{Beutler} F.,  et~al., 2011, \mn@doi [\mnras]
  {10.1111/j.1365-2966.2011.19250.x}, \href
  {https://ui.adsabs.harvard.edu/abs/2011MNRAS.416.3017B} {416, 3017}

\bibitem[\protect\citeauthoryear{{Beutler} et~al.,}{{Beutler}
  et~al.}{2012}]{Beutler2012:1204.4725}
{Beutler} F.,  et~al., 2012, \mn@doi [\mnras]
  {10.1111/j.1365-2966.2012.21136.x}, \href
  {https://ui.adsabs.harvard.edu/abs/2012MNRAS.423.3430B} {423, 3430}

\bibitem[\protect\citeauthoryear{{Blake} et~al.,}{{Blake}
  et~al.}{2011}]{Blake2011:1104.2948}
{Blake} C.,  et~al., 2011, \mn@doi [\mnras] {10.1111/j.1365-2966.2011.18903.x},
  \href {https://ui.adsabs.harvard.edu/abs/2011MNRAS.415.2876B} {415, 2876}

\bibitem[\protect\citeauthoryear{{Blas}, {Lesgourgues}  \& {Tram}}{{Blas}
  et~al.}{2011}]{class:1104.2933}
{Blas} D.,  {Lesgourgues} J.,   {Tram} T.,  2011, \mn@doi [\jcap]
  {10.1088/1475-7516/2011/07/034}, \href
  {https://ui.adsabs.harvard.edu/abs/2011JCAP...07..034B} {2011, 034}

\bibitem[\protect\citeauthoryear{{Brieden}, {Gil-Mar{\'\i}n}  \&
  {Verde}}{{Brieden} et~al.}{2021}]{Brieden2021:2106.07641}
{Brieden} S.,  {Gil-Mar{\'\i}n} H.,   {Verde} L.,  2021, \mn@doi [\jcap]
  {10.1088/1475-7516/2021/12/054}, \href
  {https://ui.adsabs.harvard.edu/abs/2021JCAP...12..054B} {2021, 054}

\bibitem[\protect\citeauthoryear{{Carrasco}, {Hertzberg}  \&
  {Senatore}}{{Carrasco} et~al.}{2012}]{Carrasco2012:1206.2926}
{Carrasco} J. J.~M.,  {Hertzberg} M.~P.,   {Senatore} L.,  2012, \mn@doi
  [Journal of High Energy Physics] {10.1007/JHEP09(2012)082}, \href
  {https://ui.adsabs.harvard.edu/abs/2012JHEP...09..082C} {2012, 82}

\bibitem[\protect\citeauthoryear{{Chapman} et~al.,}{{Chapman}
  et~al.}{2022}]{Chapman2022:2106.14961}
{Chapman} M.~J.,  et~al., 2022, \mn@doi [\mnras] {10.1093/mnras/stac1923},
  \href {https://ui.adsabs.harvard.edu/abs/2022MNRAS.516..617C} {516, 617}

\bibitem[\protect\citeauthoryear{Chuang, Kitaura, Liang, Font-Ribera, Zhao,
  McDonald  \& Tao}{Chuang et~al.}{2017}]{Chuang2017}
Chuang C.-H.,  Kitaura F.-S.,  Liang Y.,  Font-Ribera A.,  Zhao C.,  McDonald
  P.,   Tao C.,  2017, \mn@doi [Physical Review D]
  {10.1103/physrevd.95.063528}, 95

\bibitem[\protect\citeauthoryear{{Cole} et~al.,}{{Cole}
  et~al.}{2005}]{Cole2005:astro-ph/0501174}
{Cole} S.,  et~al., 2005, \mn@doi [\mnras] {10.1111/j.1365-2966.2005.09318.x},
  \href {http://adsabs.harvard.edu/abs/2005MNRAS.362..505C} {362, 505}

\bibitem[\protect\citeauthoryear{Cooke, Pettini  \& Steidel}{Cooke
  et~al.}{2018}]{Cooke2018:1710.11129}
Cooke R.~J.,  Pettini M.,   Steidel C.~C.,  2018, \mn@doi [The Astrophysical
  Journal] {10.3847/1538-4357/aaab53}, 855, 102

\bibitem[\protect\citeauthoryear{{Correa}, {Paz}, {S{\'a}nchez}, {Ruiz},
  {Padilla}  \& {Angulo}}{{Correa} et~al.}{2020}]{Correa2020}
{Correa} C.~M.,  {Paz} D.~J.,  {S{\'a}nchez} A.~G.,  {Ruiz} A.~N.,  {Padilla}
  N.~D.,   {Angulo} R.~E.,  2020, \mn@doi [\mnras] {10.1093/mnras/staa3252},
  \href {https://ui.adsabs.harvard.edu/abs/2020MNRAS.tmp.3058C} {}

\bibitem[\protect\citeauthoryear{{Cuesta-Lazaro} et~al.,}{{Cuesta-Lazaro}
  et~al.}{2023}]{Cuesta-Lazaro2023:2309.16539}
{Cuesta-Lazaro} C.,  et~al., 2023, arXiv e-prints, \href
  {https://ui.adsabs.harvard.edu/abs/2023arXiv230916539C} {p. arXiv:2309.16539}

\bibitem[\protect\citeauthoryear{{DESI Collaboration} et~al.,}{{DESI
  Collaboration} et~al.}{2016}]{desi}
{DESI Collaboration} et~al., 2016, arXiv e-prints, \href
  {https://ui.adsabs.harvard.edu/abs/2016arXiv161100036D} {p. arXiv:1611.00036}

\bibitem[\protect\citeauthoryear{{Dawson} et~al.,}{{Dawson}
  et~al.}{2013}]{Dawson2013}
{Dawson} K.~S.,  et~al., 2013, \mn@doi [\aj] {10.1088/0004-6256/145/1/10},
  \href {http://adsabs.harvard.edu/abs/2013AJ....145...10D} {145, 10}

\bibitem[\protect\citeauthoryear{{Dawson} et~al.,}{{Dawson}
  et~al.}{2016}]{Dawson2016}
{Dawson} K.~S.,  et~al., 2016, \mn@doi [\aj] {10.3847/0004-6256/151/2/44},
  \href {https://ui.adsabs.harvard.edu/abs/2016AJ....151...44D} {151, 44}

\bibitem[\protect\citeauthoryear{{Dong-P{\'a}ez} et~al.,}{{Dong-P{\'a}ez}
  et~al.}{2022}]{Dong-Paez2022:2208.00540}
{Dong-P{\'a}ez} C.~A.,  et~al., 2022, \mn@doi [arXiv e-prints]
  {10.48550/arXiv.2208.00540}, \href
  {https://ui.adsabs.harvard.edu/abs/2022arXiv220800540D} {p. arXiv:2208.00540}

\bibitem[\protect\citeauthoryear{{Eisenstein} \& {Hu}}{{Eisenstein} \&
  {Hu}}{1998}]{Eisenstein1998:astro-ph/9709112}
{Eisenstein} D.~J.,  {Hu} W.,  1998, \mn@doi [\apj] {10.1086/305424}, \href
  {https://ui.adsabs.harvard.edu/abs/1998ApJ...496..605E} {496, 605}

\bibitem[\protect\citeauthoryear{{Eisenstein} et~al.,}{{Eisenstein}
  et~al.}{2005}]{Eisenstein2005}
{Eisenstein} D.~J.,  et~al., 2005, \mn@doi [\apj] {10.1086/466512}, \href
  {http://adsabs.harvard.edu/abs/2005ApJ...633..560E} {633, 560}

\bibitem[\protect\citeauthoryear{{Emas} \& {Wulandari}}{{Emas} \&
  {Wulandari}}{2019}]{Emas2019}
{Emas} N.~P.~A.~P.,  {Wulandari} H.,  2019, in Journal of Physics Conference
  Series. p. 012026, \mn@doi{10.1088/1742-6596/1231/1/012026}

\bibitem[\protect\citeauthoryear{Feldman, Kaiser  \& Peacock}{Feldman
  et~al.}{1994}]{Feldman1994:astro-ph/9304022}
Feldman H.~A.,  Kaiser N.,   Peacock J.~A.,  1994, \mn@doi [The Astrophysical
  Journal] {10.1086/174036}, 426, 23

\bibitem[\protect\citeauthoryear{Garrison, Eisenstein  \& Pinto}{Garrison
  et~al.}{2019}]{Garrison2019:1810.02916}
Garrison L.~H.,  Eisenstein D.~J.,   Pinto P.~A.,  2019, \mn@doi [Monthly
  Notices of the Royal Astronomical Society] {10.1093/mnras/stz634}, 485, 3370

\bibitem[\protect\citeauthoryear{Garrison, Eisenstein, Ferrer, Maksimova  \&
  Pinto}{Garrison et~al.}{2021}]{Garrison2021:2110.11392}
Garrison L.~H.,  Eisenstein D.~J.,  Ferrer D.,  Maksimova N.~A.,   Pinto P.~A.,
   2021, \mn@doi [Monthly Notices of the Royal Astronomical Society]
  {10.1093/mnras/stab2482}, 508, 575

\bibitem[\protect\citeauthoryear{{Gil-Mar{\'\i}n}, {Percival}, {Verde},
  {Brownstein}, {Chuang}, {Kitaura}, {Rodr{\'\i}guez-Torres}  \&
  {Olmstead}}{{Gil-Mar{\'\i}n} et~al.}{2017}]{Gil-Marin2017:1606.00439}
{Gil-Mar{\'\i}n} H.,  {Percival} W.~J.,  {Verde} L.,  {Brownstein} J.~R.,
  {Chuang} C.-H.,  {Kitaura} F.-S.,  {Rodr{\'\i}guez-Torres} S.~A.,
  {Olmstead} M.~D.,  2017, \mn@doi [\mnras] {10.1093/mnras/stw2679}, \href
  {https://ui.adsabs.harvard.edu/abs/2017MNRAS.465.1757G} {465, 1757}

\bibitem[\protect\citeauthoryear{Green et~al.,}{Green et~al.}{2012}]{roman}
Green J.,  et~al., 2012, Wide-Field InfraRed Survey Telescope (WFIRST) Final
  Report, \mn@doi{10.48550/ARXIV.1208.4012}, \url
  {https://arxiv.org/abs/1208.4012}

\bibitem[\protect\citeauthoryear{{Grieb} et~al.,}{{Grieb}
  et~al.}{2017}]{Grieb2017:1607.03143}
{Grieb} J.~N.,  et~al., 2017, \mn@doi [\mnras] {10.1093/mnras/stw3384}, \href
  {https://ui.adsabs.harvard.edu/abs/2017MNRAS.467.2085G} {467, 2085}

\bibitem[\protect\citeauthoryear{Gualdi, Gil-Mar{\'{\i}}n  \& Verde}{Gualdi
  et~al.}{2021}]{Gualdi2021}
Gualdi D.,  Gil-Mar{\'{\i}}n H.,   Verde L.,  2021, \mn@doi [Journal of
  Cosmology and Astroparticle Physics] {10.1088/1475-7516/2021/07/008}, 2021,
  008

\bibitem[\protect\citeauthoryear{{Gunn} et~al.,}{{Gunn}
  et~al.}{1998}]{Gunn1998:astro-ph/9809085}
{Gunn} J.~E.,  et~al., 1998, \mn@doi [\aj] {10.1086/300645}, \href
  {https://ui.adsabs.harvard.edu/abs/1998AJ....116.3040G} {116, 3040}

\bibitem[\protect\citeauthoryear{Gunn et~al.,}{Gunn
  et~al.}{2006}]{Gunn2006:astro-ph/0602326}
Gunn J.~E.,  et~al., 2006, \mn@doi [The Astronomical Journal] {10.1086/500975},
  131, 2332

\bibitem[\protect\citeauthoryear{{Guth} \& {Pi}}{{Guth} \&
  {Pi}}{1982}]{Guth1982}
{Guth} A.~H.,  {Pi} S.~Y.,  1982, \mn@doi [\prl] {10.1103/PhysRevLett.49.1110},
  \href {https://ui.adsabs.harvard.edu/abs/1982PhRvL..49.1110G} {49, 1110}

\bibitem[\protect\citeauthoryear{Hadzhiyska, Eisenstein, Bose, Garrison  \&
  Maksimova}{Hadzhiyska et~al.}{2021}]{Hadzhiyska2021:2110.11408}
Hadzhiyska B.,  Eisenstein D.,  Bose S.,  Garrison L.~H.,   Maksimova N.,
  2021, \mn@doi [Monthly Notices of the Royal Astronomical Society]
  {10.1093/mnras/stab2980}, 509, 501

\bibitem[\protect\citeauthoryear{{Hadzhiyska}, {Garrison}, {Eisenstein}  \&
  {Bose}}{{Hadzhiyska} et~al.}{2022}]{Hadzhiyska2022:2110.11413}
{Hadzhiyska} B.,  {Garrison} L.~H.,  {Eisenstein} D.,   {Bose} S.,  2022,
  \mn@doi [\mnras] {10.1093/mnras/stab3066}, \href
  {https://ui.adsabs.harvard.edu/abs/2022MNRAS.509.2194H} {509, 2194}

\bibitem[\protect\citeauthoryear{{Hahn}, {Scoccimarro}, {Blanton}, {Tinker}  \&
  {Rodr{\'\i}guez-Torres}}{{Hahn} et~al.}{2017}]{Hahn2017:1609.01714}
{Hahn} C.,  {Scoccimarro} R.,  {Blanton} M.~R.,  {Tinker} J.~L.,
  {Rodr{\'\i}guez-Torres} S.~A.,  2017, \mn@doi [\mnras]
  {10.1093/mnras/stx185}, \href
  {https://ui.adsabs.harvard.edu/abs/2017MNRAS.467.1940H} {467, 1940}

\bibitem[\protect\citeauthoryear{{Hawking}}{{Hawking}}{1982}]{Hawking1982}
{Hawking} S.~W.,  1982, \mn@doi [Physics Letters B]
  {10.1016/0370-2693(82)90373-2}, \href
  {https://ui.adsabs.harvard.edu/abs/1982PhLB..115..295H} {115, 295}

\bibitem[\protect\citeauthoryear{{Hou} et~al.,}{{Hou}
  et~al.}{2021}]{Hou2021:2007.08998}
{Hou} J.,  et~al., 2021, \mn@doi [\mnras] {10.1093/mnras/staa3234}, \href
  {https://ui.adsabs.harvard.edu/abs/2021MNRAS.500.1201H} {500, 1201}

\bibitem[\protect\citeauthoryear{{Hou}, {Slepian}  \& {Cahn}}{{Hou}
  et~al.}{2023a}]{Hou2023a:2206.03625}
{Hou} J.,  {Slepian} Z.,   {Cahn} R.~N.,  2023a, \mn@doi [\mnras]
  {10.1093/mnras/stad1062}, \href
  {https://ui.adsabs.harvard.edu/abs/2023MNRAS.522.5701H} {522, 5701}

\bibitem[\protect\citeauthoryear{{Hou}, {Moradinezhad Dizgah}, {Hahn}  \&
  {Massara}}{{Hou} et~al.}{2023b}]{Hou2023b:2210.12743}
{Hou} J.,  {Moradinezhad Dizgah} A.,  {Hahn} C.,   {Massara} E.,  2023b,
  \mn@doi [\jcap] {10.1088/1475-7516/2023/03/045}, \href
  {https://ui.adsabs.harvard.edu/abs/2023JCAP...03..045H} {2023, 045}

\bibitem[\protect\citeauthoryear{{Ishiyama} et~al.,}{{Ishiyama}
  et~al.}{2021}]{Ishiyama2021:2007.14720}
{Ishiyama} T.,  et~al., 2021, \mn@doi [\mnras] {10.1093/mnras/stab1755}, \href
  {https://ui.adsabs.harvard.edu/abs/2021MNRAS.506.4210I} {506, 4210}

\bibitem[\protect\citeauthoryear{{Ivanov}, {Simonovi{\'c}}  \&
  {Zaldarriaga}}{{Ivanov} et~al.}{2020}]{Ivanov2020:1909.05277}
{Ivanov} M.~M.,  {Simonovi{\'c}} M.,   {Zaldarriaga} M.,  2020, \mn@doi [\jcap]
  {10.1088/1475-7516/2020/05/042}, \href
  {https://ui.adsabs.harvard.edu/abs/2020JCAP...05..042I} {2020, 042}

\bibitem[\protect\citeauthoryear{{Jackson}}{{Jackson}}{1972}]{Jackson1972}
{Jackson} J.~C.,  1972, \mn@doi [\mnras] {10.1093/mnras/156.1.1P}, \href
  {https://ui.adsabs.harvard.edu/abs/1972MNRAS.156P...1J} {156, 1P}

\bibitem[\protect\citeauthoryear{{Kaiser}}{{Kaiser}}{1987}]{Kaiser1987}
{Kaiser} N.,  1987, \mn@doi [\mnras] {10.1093/mnras/227.1.1}, \href
  {https://ui.adsabs.harvard.edu/abs/1987MNRAS.227....1K} {227, 1}

\bibitem[\protect\citeauthoryear{{Kitaura}, {Yepes}  \& {Prada}}{{Kitaura}
  et~al.}{2014}]{Kitaura2014:1307.3285}
{Kitaura} F.~S.,  {Yepes} G.,   {Prada} F.,  2014, \mn@doi [\mnras]
  {10.1093/mnrasl/slt172}, \href
  {https://ui.adsabs.harvard.edu/abs/2014MNRAS.439L..21K} {439, L21}

\bibitem[\protect\citeauthoryear{{Kitaura}, {Gil-Mar{\'\i}n}, {Sc{\'o}ccola},
  {Chuang}, {M{\"u}ller}, {Yepes}  \& {Prada}}{{Kitaura}
  et~al.}{2015}]{Kitaura2015:1407.1236}
{Kitaura} F.-S.,  {Gil-Mar{\'\i}n} H.,  {Sc{\'o}ccola} C.~G.,  {Chuang} C.-H.,
  {M{\"u}ller} V.,  {Yepes} G.,   {Prada} F.,  2015, \mn@doi [\mnras]
  {10.1093/mnras/stv645}, \href
  {https://ui.adsabs.harvard.edu/abs/2015MNRAS.450.1836K} {450, 1836}

\bibitem[\protect\citeauthoryear{{Kitaura} et~al.,}{{Kitaura}
  et~al.}{2016}]{Kitaura2016:1509.06400}
{Kitaura} F.-S.,  et~al., 2016, \mn@doi [\mnras] {10.1093/mnras/stv2826}, \href
  {https://ui.adsabs.harvard.edu/abs/2016MNRAS.456.4156K} {456, 4156}

\bibitem[\protect\citeauthoryear{{Klypin}, {Yepes}, {Gottl{\"o}ber}, {Prada}
  \& {He{\ss}}}{{Klypin} et~al.}{2016}]{Klypin2016:1411.4001}
{Klypin} A.,  {Yepes} G.,  {Gottl{\"o}ber} S.,  {Prada} F.,   {He{\ss}} S.,
  2016, \mn@doi [\mnras] {10.1093/mnras/stw248}, \href
  {https://ui.adsabs.harvard.edu/abs/2016MNRAS.457.4340K} {457, 4340}

\bibitem[\protect\citeauthoryear{Kobayashi, Nishimichi, Takada  \&
  Takahashi}{Kobayashi et~al.}{2020}]{Kobayashi2020}
Kobayashi Y.,  Nishimichi T.,  Takada M.,   Takahashi R.,  2020, \mn@doi
  [Physical Review D] {10.1103/physrevd.101.023510}, 101

\bibitem[\protect\citeauthoryear{{Kobayashi}, {Nishimichi}, {Takada}  \&
  {Miyatake}}{{Kobayashi} et~al.}{2022}]{Kobayashi2022:2110.06969}
{Kobayashi} Y.,  {Nishimichi} T.,  {Takada} M.,   {Miyatake} H.,  2022, \mn@doi
  [\prd] {10.1103/PhysRevD.105.083517}, \href
  {https://ui.adsabs.harvard.edu/abs/2022PhRvD.105h3517K} {105, 083517}

\bibitem[\protect\citeauthoryear{Komatsu et~al.,}{Komatsu et~al.}{2011}]{WMAP7}
Komatsu E.,  et~al., 2011, \mn@doi [The Astrophysical Journal Supplement
  Series] {10.1088/0067-0049/192/2/18}, 192, 18

\bibitem[\protect\citeauthoryear{{Kosowsky} \& {Turner}}{{Kosowsky} \&
  {Turner}}{1995}]{Kosowsky1995:astro-ph/9504071}
{Kosowsky} A.,  {Turner} M.~S.,  1995, \mn@doi [\prd]
  {10.1103/PhysRevD.52.R1739}, \href
  {https://ui.adsabs.harvard.edu/abs/1995PhRvD..52.1739K} {52, R1739}

\bibitem[\protect\citeauthoryear{{Landy} \& {Szalay}}{{Landy} \&
  {Szalay}}{1993}]{Landy1993}
{Landy} S.~D.,  {Szalay} A.~S.,  1993, \mn@doi [\apj] {10.1086/172900}, \href
  {https://ui.adsabs.harvard.edu/abs/1993ApJ...412...64L} {412, 64}

\bibitem[\protect\citeauthoryear{{Lange}, {Hearin}, {Leauthaud}, {van den
  Bosch}, {Guo}  \& {DeRose}}{{Lange} et~al.}{2022}]{Lange2022:2101.12261}
{Lange} J.~U.,  {Hearin} A.~P.,  {Leauthaud} A.,  {van den Bosch} F.~C.,  {Guo}
  H.,   {DeRose} J.,  2022, \mn@doi [\mnras] {10.1093/mnras/stab3111}, \href
  {https://ui.adsabs.harvard.edu/abs/2022MNRAS.509.1779L} {509, 1779}

\bibitem[\protect\citeauthoryear{{Lange}, {Hearin}, {Leauthaud}, {van den
  Bosch}, {Xhakaj}, {Guo}, {Wechsler}  \& {DeRose}}{{Lange}
  et~al.}{2023}]{Lange2023:2301.08692}
{Lange} J.~U.,  {Hearin} A.~P.,  {Leauthaud} A.,  {van den Bosch} F.~C.,
  {Xhakaj} E.,  {Guo} H.,  {Wechsler} R.~H.,   {DeRose} J.,  2023, \mn@doi
  [\mnras] {10.1093/mnras/stad473}, \href
  {https://ui.adsabs.harvard.edu/abs/2023MNRAS.520.5373L} {520, 5373}

\bibitem[\protect\citeauthoryear{Laureijs, Amiaux, Arduini, Auguères,
  Brinchmann, Cole  et~al.}{Laureijs et~al.}{2011}]{euclid}
Laureijs R.,  Amiaux J.,  Arduini S.,  Auguères J.~L.,  Brinchmann J.,  Cole
  R.,   et~al., 2011, preprint (\mn@eprint {arXiv} {1110.3193})

\bibitem[\protect\citeauthoryear{{Lavaux} \& {Wandelt}}{{Lavaux} \&
  {Wandelt}}{2012}]{Lavaux2012:1110.0345}
{Lavaux} G.,  {Wandelt} B.~D.,  2012, \mn@doi [\apj]
  {10.1088/0004-637X/754/2/109}, \href
  {https://ui.adsabs.harvard.edu/abs/2012ApJ...754..109L} {754, 109}

\bibitem[\protect\citeauthoryear{{Lavaux}, {Jasche}  \& {Leclercq}}{{Lavaux}
  et~al.}{2019}]{Lavaux2019:1909.06396}
{Lavaux} G.,  {Jasche} J.,   {Leclercq} F.,  2019, \mn@doi [arXiv e-prints]
  {10.48550/arXiv.1909.06396}, \href
  {https://ui.adsabs.harvard.edu/abs/2019arXiv190906396L} {p. arXiv:1909.06396}

\bibitem[\protect\citeauthoryear{{Lesgourgues} \& {Pastor}}{{Lesgourgues} \&
  {Pastor}}{2014}]{Lesgourgues2014:1404.1740}
{Lesgourgues} J.,  {Pastor} S.,  2014, \mn@doi [New Journal of Physics]
  {10.1088/1367-2630/16/6/065002}, \href
  {https://ui.adsabs.harvard.edu/abs/2014NJPh...16f5002L} {16, 065002}

\bibitem[\protect\citeauthoryear{{Levi} et~al.,}{{Levi}
  et~al.}{2019}]{Levi2019}
{Levi} M.,  et~al., 2019, in Bulletin of the American Astronomical Society.
  p.~57 (\mn@eprint {arXiv} {1907.10688})

\bibitem[\protect\citeauthoryear{{Lippich} \& {S{\'a}nchez}}{{Lippich} \&
  {S{\'a}nchez}}{2021}]{Lippich2021:2012.08529}
{Lippich} M.,  {S{\'a}nchez} A.~G.,  2021, \mn@doi [\mnras]
  {10.1093/mnras/stab2820}, \href
  {https://ui.adsabs.harvard.edu/abs/2021MNRAS.508.3771L} {508, 3771}

\bibitem[\protect\citeauthoryear{Maksimova, Garrison, Eisenstein, Hadzhiyska,
  Bose  \& Satterthwaite}{Maksimova et~al.}{2021}]{Maksimova2021:2110.11398}
Maksimova N.~A.,  Garrison L.~H.,  Eisenstein D.~J.,  Hadzhiyska B.,  Bose S.,
   Satterthwaite T.~P.,  2021, \mn@doi [Monthly Notices of the Royal
  Astronomical Society] {10.1093/mnras/stab2484}, 508, 4017

\bibitem[\protect\citeauthoryear{{Maraston} et~al.,}{{Maraston}
  et~al.}{2013}]{Maraston2013:1207.6114}
{Maraston} C.,  et~al., 2013, \mn@doi [\mnras] {10.1093/mnras/stt1424}, \href
  {https://ui.adsabs.harvard.edu/abs/2013MNRAS.435.2764M} {435, 2764}

\bibitem[\protect\citeauthoryear{{Massara} et~al.,}{{Massara}
  et~al.}{2022}]{Massara2022}
{Massara} E.,  et~al., 2022, arXiv e-prints, \href
  {https://ui.adsabs.harvard.edu/abs/2022arXiv220601709M} {p. arXiv:2206.01709}

\bibitem[\protect\citeauthoryear{{Moon} et~al.,}{{Moon}
  et~al.}{2023}]{Moon2023:2304.08427}
{Moon} J.,  et~al., 2023, \mn@doi [arXiv e-prints] {10.48550/arXiv.2304.08427},
  \href {https://ui.adsabs.harvard.edu/abs/2023arXiv230408427M} {p.
  arXiv:2304.08427}

\bibitem[\protect\citeauthoryear{{Moradinezhad Dizgah}, {Biagetti},
  {Sefusatti}, {Desjacques}  \& {Nore{\~n}a}}{{Moradinezhad Dizgah}
  et~al.}{2021}]{Moradinezhad2021:2010.14523}
{Moradinezhad Dizgah} A.,  {Biagetti} M.,  {Sefusatti} E.,  {Desjacques} V.,
  {Nore{\~n}a} J.,  2021, \mn@doi [\jcap] {10.1088/1475-7516/2021/05/015},
  \href {https://ui.adsabs.harvard.edu/abs/2021JCAP...05..015M} {2021, 015}

\bibitem[\protect\citeauthoryear{{Mukhanov}}{{Mukhanov}}{2007}]{Mukhanov2007}
{Mukhanov} S.,  2007, \mn@doi [Journal of Physics A Mathematical General]
  {10.1088/1751-8113/40/25/S01}, \href
  {https://ui.adsabs.harvard.edu/abs/2007JPhA...40.6561M} {40, 6561}

\bibitem[\protect\citeauthoryear{{Nadathur}, {Carter}  \&
  {Percival}}{{Nadathur} et~al.}{2019}]{Nadathur2019b}
{Nadathur} S.,  {Carter} P.,   {Percival} W.~J.,  2019, \mn@doi [\mnras]
  {10.1093/mnras/sty2799}, \href
  {https://ui.adsabs.harvard.edu/abs/2019MNRAS.482.2459N} {482, 2459}

\bibitem[\protect\citeauthoryear{{Oogi} et~al.,}{{Oogi}
  et~al.}{2023}]{Oogi2023:2207.14689}
{Oogi} T.,  et~al., 2023, \mn@doi [\mnras] {10.1093/mnras/stad2401}, \href
  {https://ui.adsabs.harvard.edu/abs/2023MNRAS.525.3879O} {525, 3879}

\bibitem[\protect\citeauthoryear{Paillas, Cai, Padilla  \& Sánchez}{Paillas
  et~al.}{2021}]{Paillas2021}
Paillas E.,  Cai Y.-C.,  Padilla N.,   Sánchez A.~G.,  2021, \mn@doi [Monthly
  Notices of the Royal Astronomical Society] {10.1093/mnras/stab1654}, 505,
  5731–5752

\bibitem[\protect\citeauthoryear{{Paillas} et~al.,}{{Paillas}
  et~al.}{2023}]{Paillas2022:2209.04310}
{Paillas} E.,  et~al., 2023, \mn@doi [\mnras] {10.1093/mnras/stad1017}, \href
  {https://ui.adsabs.harvard.edu/abs/2023MNRAS.522..606P} {522, 606}

\bibitem[\protect\citeauthoryear{{Peebles}}{{Peebles}}{1980}]{Peebles1980}
{Peebles} P.~J.~E.,  1980, {The large-scale structure of the universe}

\bibitem[\protect\citeauthoryear{{Peebles} \& {Yu}}{{Peebles} \&
  {Yu}}{1970}]{Peebles1970}
{Peebles} P.~J.~E.,  {Yu} J.~T.,  1970, \mn@doi [\apj] {10.1086/150713}, \href
  {https://ui.adsabs.harvard.edu/abs/1970ApJ...162..815P} {162, 815}

\bibitem[\protect\citeauthoryear{{Percival} et~al.,}{{Percival}
  et~al.}{2001}]{Percival2001:astro-ph/0105252}
{Percival} W.~J.,  et~al., 2001, \mn@doi [\mnras]
  {10.1046/j.1365-8711.2001.04827.x}, \href
  {https://ui.adsabs.harvard.edu/abs/2001MNRAS.327.1297P} {327, 1297}

\bibitem[\protect\citeauthoryear{{Percival} et~al.,}{{Percival}
  et~al.}{2002}]{Percival2002:astro-ph/0206256}
{Percival} W.~J.,  et~al., 2002, \mn@doi [\mnras]
  {10.1046/j.1365-8711.2002.06001.x}, \href
  {https://ui.adsabs.harvard.edu/abs/2002MNRAS.337.1068P} {337, 1068}

\bibitem[\protect\citeauthoryear{{Percival}, {Friedrich}, {Sellentin}  \&
  {Heavens}}{{Percival} et~al.}{2022}]{Percival2022}
{Percival} W.~J.,  {Friedrich} O.,  {Sellentin} E.,   {Heavens} A.,  2022,
  \mn@doi [\mnras] {10.1093/mnras/stab3540}, \href
  {https://ui.adsabs.harvard.edu/abs/2022MNRAS.510.3207P} {510, 3207}

\bibitem[\protect\citeauthoryear{{Philcox}}{{Philcox}}{2022}]{Philcox2022:2206.04227}
{Philcox} O. H.~E.,  2022, \mn@doi [\prd] {10.1103/PhysRevD.106.063501}, \href
  {https://ui.adsabs.harvard.edu/abs/2022PhRvD.106f3501P} {106, 063501}

\bibitem[\protect\citeauthoryear{{Philcox} \& {Ivanov}}{{Philcox} \&
  {Ivanov}}{2022}]{Philcox2021a:2112.04515}
{Philcox} O. H.~E.,  {Ivanov} M.~M.,  2022, \mn@doi [\prd]
  {10.1103/PhysRevD.105.043517}, \href
  {https://ui.adsabs.harvard.edu/abs/2022PhRvD.105d3517P} {105, 043517}

\bibitem[\protect\citeauthoryear{{Philcox}, {Hou}  \& {Slepian}}{{Philcox}
  et~al.}{2021}]{Philcox2021b:2108.01670}
{Philcox} O. H.~E.,  {Hou} J.,   {Slepian} Z.,  2021, arXiv e-prints, \href
  {https://ui.adsabs.harvard.edu/abs/2021arXiv210801670P} {p. arXiv:2108.01670}

\bibitem[\protect\citeauthoryear{{Planck Collaboration} et~al.,}{{Planck
  Collaboration} et~al.}{2014}]{Planck2014}
{Planck Collaboration} et~al., 2014, \mn@doi [\aap]
  {10.1051/0004-6361/201321591}, \href
  {https://ui.adsabs.harvard.edu/abs/2014A&A...571A..16P} {571, A16}

\bibitem[\protect\citeauthoryear{{Planck Collaboration} et~al.,}{{Planck
  Collaboration} et~al.}{2020}]{Planck2020}
{Planck Collaboration} et~al., 2020, \mn@doi [\aap]
  {10.1051/0004-6361/201833910}, \href
  {https://ui.adsabs.harvard.edu/abs/2020A&A...641A...6P} {641, A6}

\bibitem[\protect\citeauthoryear{{Prada}, {Behroozi}, {Ishiyama}, {Klypin}  \&
  {P{\'e}rez}}{{Prada} et~al.}{2023}]{Prada2023:2304.11911}
{Prada} F.,  {Behroozi} P.,  {Ishiyama} T.,  {Klypin} A.,   {P{\'e}rez} E.,
  2023, \mn@doi [arXiv e-prints] {10.48550/arXiv.2304.11911}, \href
  {https://ui.adsabs.harvard.edu/abs/2023arXiv230411911P} {p. arXiv:2304.11911}

\bibitem[\protect\citeauthoryear{{Reid} et~al.,}{{Reid}
  et~al.}{2012}]{Reid2012:1203.6641}
{Reid} B.~A.,  et~al., 2012, \mn@doi [\mnras]
  {10.1111/j.1365-2966.2012.21779.x}, \href
  {https://ui.adsabs.harvard.edu/abs/2012MNRAS.426.2719R} {426, 2719}

\bibitem[\protect\citeauthoryear{{Reid} et~al.,}{{Reid}
  et~al.}{2016}]{Reid2016:1509.06529}
{Reid} B.,  et~al., 2016, \mn@doi [\mnras] {10.1093/mnras/stv2382}, \href
  {https://ui.adsabs.harvard.edu/abs/2016MNRAS.455.1553R} {455, 1553}

\bibitem[\protect\citeauthoryear{{Riess} et~al.,}{{Riess}
  et~al.}{2018}]{Riess:2018:1801.01120}
{Riess} A.~G.,  et~al., 2018, \mn@doi [\apj] {10.3847/1538-4357/aaadb7}, \href
  {https://ui.adsabs.harvard.edu/abs/2018ApJ...855..136R} {855, 136}

\bibitem[\protect\citeauthoryear{{Ross}, {Samushia}, {Howlett}, {Percival},
  {Burden}  \& {Manera}}{{Ross} et~al.}{2015}]{Ross2015:1409.3242}
{Ross} A.~J.,  {Samushia} L.,  {Howlett} C.,  {Percival} W.~J.,  {Burden} A.,
  {Manera} M.,  2015, \mn@doi [\mnras] {10.1093/mnras/stv154}, \href
  {https://ui.adsabs.harvard.edu/abs/2015MNRAS.449..835R} {449, 835}

\bibitem[\protect\citeauthoryear{Ross et~al.,}{Ross
  et~al.}{2016}]{Ross2016:1607.03145}
Ross A.~J.,  et~al., 2016, \mn@doi [Monthly Notices of the Royal Astronomical
  Society] {10.1093/mnras/stw2372}, 464, 1168

\bibitem[\protect\citeauthoryear{{S{\'a}nchez}}{{S{\'a}nchez}}{2020}]{Sanchez2020}
{S{\'a}nchez} A.~G.,  2020, \mn@doi [\prd] {10.1103/PhysRevD.102.123511}, \href
  {https://ui.adsabs.harvard.edu/abs/2020PhRvD.102l3511S} {102, 123511}

\bibitem[\protect\citeauthoryear{{S{\'a}nchez} et~al.,}{{S{\'a}nchez}
  et~al.}{2017}]{Sanchez2016}
{S{\'a}nchez} A.~G.,  et~al., 2017, \mn@doi [\mnras] {10.1093/mnras/stw2443},
  \href {https://ui.adsabs.harvard.edu/abs/2017MNRAS.464.1640S} {464, 1640}

\bibitem[\protect\citeauthoryear{{Satpathy} et~al.,}{{Satpathy}
  et~al.}{2017}]{Satpathy2017:1607.03148}
{Satpathy} S.,  et~al., 2017, \mn@doi [\mnras] {10.1093/mnras/stx883}, \href
  {https://ui.adsabs.harvard.edu/abs/2017MNRAS.469.1369S} {469, 1369}

\bibitem[\protect\citeauthoryear{Schöneberg, Lesgourgues  \&
  Hooper}{Schöneberg et~al.}{2019}]{Schoneberg2019:1907.11594}
Schöneberg N.,  Lesgourgues J.,   Hooper D.~C.,  2019, \mn@doi [Journal of
  Cosmology and Astroparticle Physics] {10.1088/1475-7516/2019/10/029}, 2019,
  029

\bibitem[\protect\citeauthoryear{{Scolnic} et~al.,}{{Scolnic}
  et~al.}{2015}]{Scolnic2015:1508.05361}
{Scolnic} D.,  et~al., 2015, \mn@doi [\apj] {10.1088/0004-637X/815/2/117},
  \href {https://ui.adsabs.harvard.edu/abs/2015ApJ...815..117S} {815, 117}

\bibitem[\protect\citeauthoryear{{Semenaite} et~al.,}{{Semenaite}
  et~al.}{2022}]{Semenaite2022:2111.0315}
{Semenaite} A.,  et~al., 2022, \mn@doi [\mnras] {10.1093/mnras/stac829}, \href
  {https://ui.adsabs.harvard.edu/abs/2022MNRAS.512.5657S} {512, 5657}

\bibitem[\protect\citeauthoryear{{Semenaite} et~al.,}{{Semenaite}
  et~al.}{2023}]{Semenaite2023:2210.07304}
{Semenaite} A.,  et~al., 2023, \mn@doi [\mnras] {10.1093/mnras/stad849}, \href
  {https://ui.adsabs.harvard.edu/abs/2023MNRAS.521.5013S} {521, 5013}

\bibitem[\protect\citeauthoryear{{Sinha} \& {Garrison}}{{Sinha} \&
  {Garrison}}{2020}]{corrfunc}
{Sinha} M.,  {Garrison} L.~H.,  2020, \mn@doi [\mnras] {10.1093/mnras/stz3157},
  \href {https://ui.adsabs.harvard.edu/abs/2020MNRAS.491.3022S} {491, 3022}

\bibitem[\protect\citeauthoryear{{Slepian} et~al.,}{{Slepian}
  et~al.}{2017}]{Slepian2017a:1512.02231}
{Slepian} Z.,  et~al., 2017, \mn@doi [\mnras] {10.1093/mnras/stw3234}, \href
  {https://ui.adsabs.harvard.edu/abs/2017MNRAS.468.1070S} {468, 1070}

\bibitem[\protect\citeauthoryear{{Speagle}}{{Speagle}}{2020}]{dynesty:1904.02180}
{Speagle} J.~S.,  2020, \mn@doi [\mnras] {10.1093/mnras/staa278}, \href
  {https://ui.adsabs.harvard.edu/abs/2020MNRAS.493.3132S} {493, 3132}

\bibitem[\protect\citeauthoryear{{Sugiyama} et~al.,}{{Sugiyama}
  et~al.}{2023}]{Sugiyama2023:2302.06808}
{Sugiyama} N.~S.,  et~al., 2023, \mn@doi [\mnras] {10.1093/mnras/stad1505},
  \href {https://ui.adsabs.harvard.edu/abs/2023MNRAS.523.3133S} {523, 3133}

\bibitem[\protect\citeauthoryear{{Sunyaev} \& {Zeldovich}}{{Sunyaev} \&
  {Zeldovich}}{1970}]{Sunyaev1970}
{Sunyaev} R.~A.,  {Zeldovich} Y.~B.,  1970, \mn@doi [\apss]
  {10.1007/BF00653471}, \href
  {https://ui.adsabs.harvard.edu/abs/1970Ap&SS...7....3S} {7, 3}

\bibitem[\protect\citeauthoryear{{Tr{\"o}ster} et~al.,}{{Tr{\"o}ster}
  et~al.}{2020}]{Troster2020:1909.11006}
{Tr{\"o}ster} T.,  et~al., 2020, \mn@doi [\aap] {10.1051/0004-6361/201936772},
  \href {https://ui.adsabs.harvard.edu/abs/2020A&A...633L..10T} {633, L10}

\bibitem[\protect\citeauthoryear{{Valogiannis} \& {Dvorkin}}{{Valogiannis} \&
  {Dvorkin}}{2022a}]{Valogiannis2021}
{Valogiannis} G.,  {Dvorkin} C.,  2022a, \mn@doi [\prd]
  {10.1103/PhysRevD.105.103534}, \href
  {https://ui.adsabs.harvard.edu/abs/2022PhRvD.105j3534V} {105, 103534}

\bibitem[\protect\citeauthoryear{{Valogiannis} \& {Dvorkin}}{{Valogiannis} \&
  {Dvorkin}}{2022b}]{Valogiannis2022:2204.13717}
{Valogiannis} G.,  {Dvorkin} C.,  2022b, \mn@doi [\prd]
  {10.1103/PhysRevD.106.103509}, \href
  {https://ui.adsabs.harvard.edu/abs/2022PhRvD.106j3509V} {106, 103509}

\bibitem[\protect\citeauthoryear{{White}}{{White}}{2016}]{White2016:1609.08632}
{White} M.,  2016, \mn@doi [\jcap] {10.1088/1475-7516/2016/11/057}, \href
  {https://ui.adsabs.harvard.edu/abs/2016JCAP...11..057W} {2016, 057}

\bibitem[\protect\citeauthoryear{{York} et~al.,}{{York}
  et~al.}{2000}]{York2000:astro-ph/0006396}
{York} D.~G.,  et~al., 2000, \mn@doi [\aj] {10.1086/301513}, \href
  {https://ui.adsabs.harvard.edu/abs/2000AJ....120.1579Y} {120, 1579}

\bibitem[\protect\citeauthoryear{{Yu}, {Seljak}, {Li}  \& {Singh}}{{Yu}
  et~al.}{2023}]{Yu2023:2211.16794}
{Yu} B.,  {Seljak} U.,  {Li} Y.,   {Singh} S.,  2023, \mn@doi [\jcap]
  {10.1088/1475-7516/2023/04/057}, \href
  {https://ui.adsabs.harvard.edu/abs/2023JCAP...04..057Y} {2023, 057}

\bibitem[\protect\citeauthoryear{{Yuan}, {Garrison}, {Hadzhiyska}, {Bose}  \&
  {Eisenstein}}{{Yuan} et~al.}{2022a}]{Yuan2022:2110.11412}
{Yuan} S.,  {Garrison} L.~H.,  {Hadzhiyska} B.,  {Bose} S.,   {Eisenstein}
  D.~J.,  2022a, \mn@doi [\mnras] {10.1093/mnras/stab3355}, \href
  {https://ui.adsabs.harvard.edu/abs/2022MNRAS.510.3301Y} {510, 3301}

\bibitem[\protect\citeauthoryear{{Yuan}, {Garrison}, {Eisenstein}  \&
  {Wechsler}}{{Yuan} et~al.}{2022b}]{Yuan2022b:2203.11963}
{Yuan} S.,  {Garrison} L.~H.,  {Eisenstein} D.~J.,   {Wechsler} R.~H.,  2022b,
  \mn@doi [\mnras] {10.1093/mnras/stac1830}, \href
  {https://ui.adsabs.harvard.edu/abs/2022MNRAS.515..871Y} {515, 871}

\bibitem[\protect\citeauthoryear{{Zhai} et~al.,}{{Zhai}
  et~al.}{2023}]{Zhai2023:2203.08999}
{Zhai} Z.,  et~al., 2023, \mn@doi [\apj] {10.3847/1538-4357/acc65b}, \href
  {https://ui.adsabs.harvard.edu/abs/2023ApJ...948...99Z} {948, 99}

\bibitem[\protect\citeauthoryear{{Zhang} \& {Cai}}{{Zhang} \&
  {Cai}}{2022}]{Zhang2022:2111.05739}
{Zhang} P.,  {Cai} Y.,  2022, \mn@doi [\jcap] {10.1088/1475-7516/2022/01/031},
  \href {https://ui.adsabs.harvard.edu/abs/2022JCAP...01..031Z} {2022, 031}

\bibitem[\protect\citeauthoryear{{Zheng}, {Coil}  \& {Zehavi}}{{Zheng}
  et~al.}{2007}]{Zheng2007}
{Zheng} Z.,  {Coil} A.~L.,   {Zehavi} I.,  2007, \mn@doi [\apj]
  {10.1086/521074}, \href {http://adsabs.harvard.edu/abs/2007ApJ...667..760Z}
  {667, 760}

\bibitem[\protect\citeauthoryear{{d'Amico}, {Gleyzes}, {Kokron}, {Markovic},
  {Senatore}, {Zhang}, {Beutler}  \& {Gil-Mar{\'\i}n}}{{d'Amico}
  et~al.}{2020}]{d'Amico2020:1909.05271}
{d'Amico} G.,  {Gleyzes} J.,  {Kokron} N.,  {Markovic} K.,  {Senatore} L.,
  {Zhang} P.,  {Beutler} F.,   {Gil-Mar{\'\i}n} H.,  2020, \mn@doi [\jcap]
  {10.1088/1475-7516/2020/05/005}, \href
  {https://ui.adsabs.harvard.edu/abs/2020JCAP...05..005D} {2020, 005}

\bibitem[\protect\citeauthoryear{{de Mattia} et~al.,}{{de Mattia}
  et~al.}{2021}]{deMattia2021:2007.09008}
{de Mattia} A.,  et~al., 2021, \mn@doi [\mnras] {10.1093/mnras/staa3891}, \href
  {https://ui.adsabs.harvard.edu/abs/2021MNRAS.501.5616D} {501, 5616}

\bibitem[\protect\citeauthoryear{{de Salas} \& {Pastor}}{{de Salas} \&
  {Pastor}}{2016}]{deSalas2016:1606.06986}
{de Salas} P.~F.,  {Pastor} S.,  2016, \mn@doi [\jcap]
  {10.1088/1475-7516/2016/07/051}, \href
  {https://ui.adsabs.harvard.edu/abs/2016JCAP...07..051D} {2016, 051}

\bibitem[\protect\citeauthoryear{{eBOSS Collaboration} et~al.,}{{eBOSS
  Collaboration} et~al.}{2020}]{eboss2020}
{eBOSS Collaboration} et~al., 2020, arXiv e-prints, \href
  {https://ui.adsabs.harvard.edu/abs/2020arXiv200708991E} {p. arXiv:2007.08991}

\makeatother
\end{thebibliography}



\appendix

\section{Galaxy-halo connection constraints}
\label{ap:hod_constraints}

\begin{table}
    \renewcommand{\arraystretch}{1.4}
    \centering
    \caption{HOD parameter constraints for the base-$\Lambda$CDM fits on the CMASS galaxy catalogue, using the combination of density-split clustering and the galaxy 2PCF, or using only the latter. In those cases where there is no constraining power other than that from the prior, we use an em dash.
    }
    \rowcolors{2}{gray!15}{white}
    \begin{tabular} {|c  | c  c  | c   c|}
    \hline
    & \multicolumn{2}{c|}{density-split + galaxy 2PCF} & \multicolumn{2}{c|}{galaxy 2PCF} \\
     \hline
     Parameter &  best-fit & mean $\pm \sigma$ &  best-fit & mean $\pm \sigma$ \\
    \hline
    {$\log M_1       $} & $13.6662$ & $13.74^{+0.10}_{-0.20}$ & 13.9449 & $> 13.9     $\\
    {$\log M_{\rm cut}$} & $12.6617$ & $12.698^{+0.074}_{-0.16}$ & 12.7406 & $12.71^{+0.13}_{-0.18}     $\\
    {$\alpha         $} & $1.3452$ & $1.18^{+0.24}_{-0.15}$ & 1.4275 & $1.08^{+0.20}_{-0.26}$ \\
    {$\alpha_{\rm vel, s}$} & $0.7555$ & $< 0.857$ & 0.8682 & $0.963^{+0.093}_{-0.25}$\\
    {$\alpha_{\rm vel, c}$} & $0.0079$ & $< 0.147$ & 0.0856 & ---- \\
    {$\log \sigma    $} & $-0.5054$ & $-0.613^{+0.16}_{-0.013}$ & -1.0385 & $< -1.24$\\
    {$\kappa         $} & $0.0312$ & $< 0.524$ & 0.6525 & > 0.622\\
    {$B_{\rm cen}    $} & $-0.4359$ & $-0.306^{+0.040}_{-0.18}$ & 0.0119 & $-0.20^{+0.12}_{-0.24}$\\
    {$B_{\rm sat}    $} & $-0.2179$ & $-0.39\pm 0.30$ & -0.4332 & $-0.18^{+0.28}_{-0.76}$\\
    \hline
    \end{tabular}
    \label{tab:hod_constraints}
\end{table}

\begin{figure*}
    \centering
    \includegraphics[width=\textwidth]{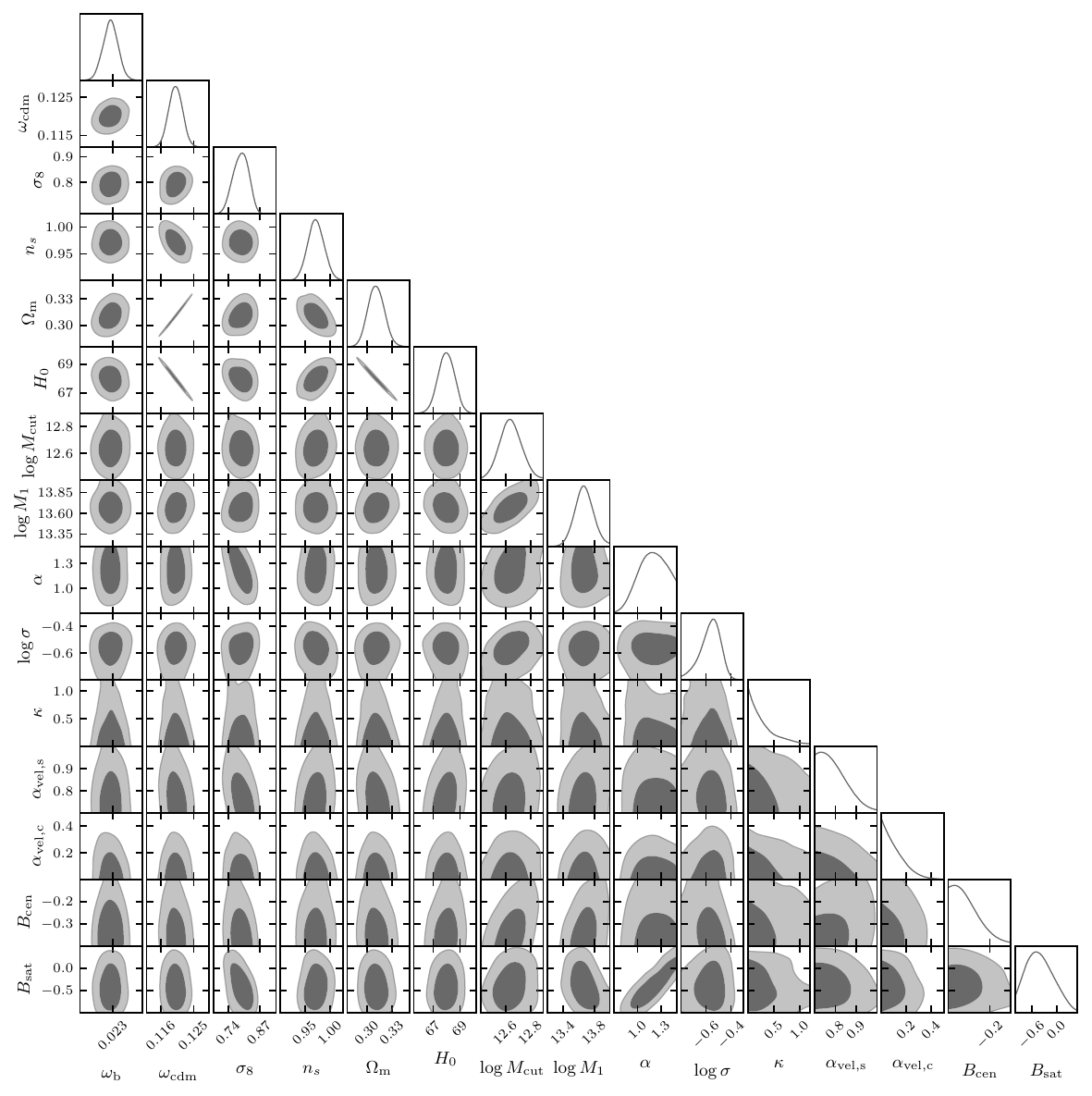}
    \caption{Similar to Fig.~\ref{fig:cosmo_inference_baseline}, but showing the full posterior distribution of cosmological and HOD parameters from the base-$\Lambda$CDM fits with our baseline configuration. \href{https://github.com/florpi/sunbird/blob/main/paper_figures/boss/cosmo_inferece_cmass_full_posterior.py}{\faGithub}}
    \label{fig:base_lcdm_full_posterior}
\end{figure*}

Figure~\ref{fig:base_lcdm_full_posterior} shows the full posterior distribution, including HOD parameters, for the base-$\Lambda$CDM fit using the baseline configuration of our analysis, where we combine the DSC and galaxy 2PCF multipoles. The constraints are largely consistent with CMASS HOD constraints with the redshift-space 2PCF in \cite{Yuan2022:2110.11412} and the cosmology-marginalized constraints in \cite{Yuan2022b:2203.11963}. In terms of the vanilla parameters, both \cite{Yuan2022:2110.11412} and \cite{Yuan2022b:2203.11963} found the best fit to favor $\log M_\mathrm{cut} \approx 12.8$, $\log M_1 \approx 14.0$, and $\alpha \approx 1.0$, all in excellent agreement with our posterior constraints here. However, \cite{Yuan2022:2110.11412} and \cite{Yuan2022b:2203.11963} found a lower $\log \sigma$, which is consistent with the 2PCF-only constraints in Table~\ref{tab:hod_constraints}, but significantly less than the DSC + 2PCF results. In terms of velocity biases, we find a lower central and satellite velocity bias compared to the two previous studies.

In terms of galaxy assembly bias, we find negative $B_\mathrm{cen}$ and $B_\mathrm{sat}$, which is qualitatively consistent with \cite{Yuan2022:2110.11412}. However, the amplitude of the inferred central assembly bias is significantly larger than that of \cite{Yuan2022:2110.11412} and \cite{Yuan2022b:2203.11963}. Physically, negative central assembly bias means that central galaxies prefer denser environments, consistent with our intuition of large red galaxies preferentially occupying highly biased cluster environments. \cite{Yuan2022b:2203.11963} also found a mild degeneracy between environment-based assembly bias and velocity bias. By allowing galaxies to preferentially occupy halos in denser environments, these galaxies would occupy deeper potential wells and thus have higher peculiar velocities, reducing the need to invoke additional velocity bias. This could potentially explain the lower inferred velocity bias we find.

We also notice that even though our priors on HOD parameters are fairly broad, the marginalized posteriors for some parameters hit the lower bound of the prior limits. This is the case for the velocity bias of satellites $\alpha_{\rm vel, s}$, and the velocity bias of centrals $B_{\rm cen}$. Therefore, we advise the reader to interpret our results for these parameters as trends rather than reliable central values, error bars, or upper limits. However, our cosmology results are likely not significantly impacted, as there is weak to no correlation between these parameters and cosmological parameters. We plan to further explore different choices of priors on the HOD parameters in our upcoming work.

\section{Dependence on the model parameters}

\begin{figure*}
    \centering
    \includegraphics[width=0.9\textwidth]{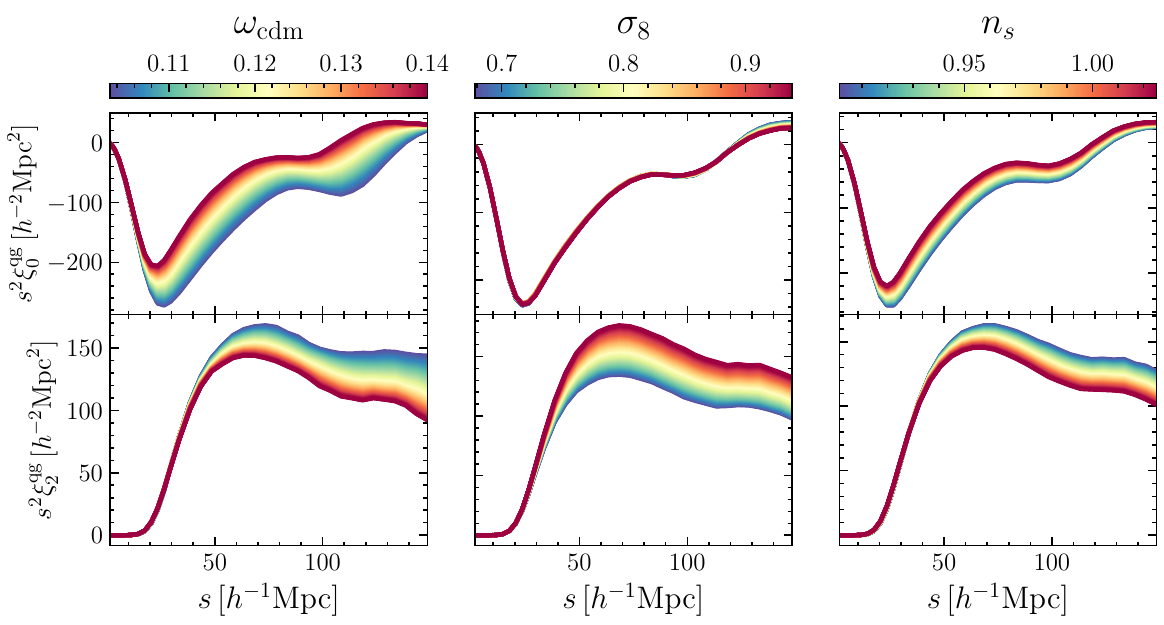}
    \caption{The dependence on model parameters as predicted by our emulator, for the cross-correlation function between the most underdense quintile ($\Q_0$) and the galaxy field. Each subpanel shows the monopole and quadrupole moments, rescaled by $s^2$ to highlight the model features at large scales. The ranges that are plotted for each parameter correspond to the prior ranges used in the likelihood analysis. \href{https://github.com/florpi/sunbird/blob/main/paper_figures/boss/parameter_dependence.py}{\faGithub}}
    \label{fig:parameter_dependence}
\end{figure*}

To explore the source of the cosmological constraining power of DSC, Fig.~\ref{fig:parameter_dependence} illustrates how the model vector responds to changes in different cosmological parameters. For brevity, we only show the cross-correlation function between the first quintile, $\Q_0$ and the galaxy field, but we have verified that the trends for other quintiles are similar in nature.

Increasing $\omegac$ results in a decrease in the absolute value of the amplitude of the monopole and quadrupole. There are several effects that come into play here. First, changing $\omegac$ has an effect on the matter clustering, which affects the shape of the galaxy density PDF, and consequently the identification of density quintiles. $\omegac$ is also tightly correlated with $\Omegam$, which determines $f\sigma_8$, which in turn sets the amplitude of velocity fluctuations and affects RSD. Moreover, the changes in $\omegac$ produce AP distortions, which affect the quintile-galaxy pair separations and become more severe the more it deviates from the fiducial cosmology used to convert redshift to distances. AP distortions not only change the amplitude of the multipoles, but also produce horizontal shifts in the monopole, changing the scale of the acoustic feature.

The amplitude of the quadrupole correlates positively with $\sigma_8$, which is in agreement with our expectations of linear theory regarding how the amplitude of velocity fluctuations scales with this parameter. Interestingly, the monopole amplitude remains fixed regardless of the value of $\sigma_8$. This is a consequence of the way in which we populate dark matter halos with galaxies: for those HOD parameter combinations where the resulting galaxy number density is greater than the CMASS number density, $\approx 3.5 \times 10^{-4} (\hMpc)^3$, the catalogues are downsampled to match the target. This effectively fixes $b\sigma_8$ in our training sample, where $b$ is the linear galaxy bias. As we vary $\sigma_8$, the galaxy bias also changes, keeping the amplitude of the monopole fixed.

Finally, we see that the spectral index of the primordial power spectrum, $n_s$, is anti-correlated with the amplitude of the monopole and quadrupole, but does not produce horizontal shifts in the profiles as in AP distortions.


\bsp	
\label{lastpage}
\end{document}